\documentclass[%
 reprint,
superscriptaddress,
 amsmath,amssymb,
 aps,
]{revtex4-2}
\usepackage[version=3]{mhchem} 
\usepackage{amsmath}
\usepackage{amssymb}
\usepackage{siunitx}
\usepackage{graphicx}
\usepackage{dcolumn}
\usepackage{bm}

\begin{document}


\title{Efficient Analysis of Photoluminescence Images for the Classification \\ of Single-Photon Emitters}

\author{Leah R. Narun}
\affiliation{%
 Quantum Engineering Laboratory, Department of Electrical and Systems Engineering, University of Pennsylvania, 200 South 33rd Street, Philadelphia, PA, 19104, USA
}%
\affiliation{Department of Materials Science and Engineering, University of Pennsylvania, 3231 Walnut Street Philadelphia, Pennsylvania 19104, USA}
\author{Rebecca E. K. Fishman}
\affiliation{%
 Quantum Engineering Laboratory, Department of Electrical and Systems Engineering, University of Pennsylvania, 200 South 33rd Street, Philadelphia, PA, 19104, USA
}%
\affiliation{Department of Physics and Astronomy, University of Pennsylvania,209 S. 33rd St. Philadelphia, Pennsylvania 19104, USA}
\author{Henry J. Shulevitz}
\author{Raj N. Patel}
\author{Lee C. Bassett}
\email[Corresponding Author. Email: ]{lbassett@seas.upenn.edu}
\affiliation{%
 Quantum Engineering Laboratory, Department of Electrical and Systems Engineering, University of Pennsylvania, 200 South 33rd Street, Philadelphia, PA, 19104, USA
}%

\date{\today}

\begin{abstract}
Solid-state single-photon emitters (SPE) are a basis for emerging technologies such as quantum communication and quantum sensing.
SPE based on fluorescent point defects are ubiquitous in semiconductors and insulators, and new systems with desirable properties for quantum information science may exist amongst the vast number of unexplored defects.
However, the characterization of new SPE typically relies on time-consuming techniques for identifying point source emitters by eye in photoluminescence (PL) images. 
This manual strategy is a bottleneck for discovering new SPE, motivating a more efficient method for characterizing emitters in PL images.
Here we present a quantitative method using image analysis and regression fitting to automatically identify Gaussian emitters in PL images and classify them according to their stability, shape, and intensity relative to the background. 
We demonstrate efficient emitter classification for SPEs in nanodiamond arrays and hexagonal boron nitride flakes. 
Adaptive criteria detect SPE in both samples despite variation in emitter intensity, stability, and background features. 
The detection criteria can be tuned for specific material systems and experimental setups to accommodate the diverse properties of SPE.
\end{abstract}

\maketitle


\section{Introduction}
Solid-state single-photon emitters (SPE) are a promising basis for future quantum technologies such as quantum computing \cite{Childress2013, Liu2019, Jiang2007, Atature2018}, sensing \cite{Degen2017, Schirhagl2014}, memory \cite{Heshami2016, Bradley2019}, and communication \cite{Gao2015,Wehner2018}.
Notable solid-state SPE systems include quantum dots \cite{Arakawa2020, Somaschi2016, Loredo2016} and fluorescent atomic defects within wide band-gap semiconductors such as diamond and silicon carbide \cite{Aharonovich2016, Wolfowicz2021}.
More recently, interest has focused on two-dimensional (2D) materials such as hexagonal boron nitride (h-BN) \cite{Tran2016,Koperski2018, Exarhos2019, Stern2019,Reserbat-Plantey2021, Azzam2021} and transition-metal dichalcogenides \cite{Palacios-Berraquero2017,Chakraborty2019, Dang2020}.  
There is no known ideal SPE system for all quantum applications, and even those that are suited for particular applications still come with trade-offs in their optical or material properties \cite{Bassett2019}. 
Improved SPE may emerge from the engineering of known systems or the discovery of new defects predicted by machine learning and ab initio calculations \cite{Shenoy2020,Ferrenti2020}.
Many materials are mostly unexplored for SPE or in the initial stages of SPE investigation, including zinc sulfide \cite{Stewart2019}, zinc oxide \cite{Linpeng2018}, titanium dioxide \cite{Chung2018}, gallium nitride \cite{Berhane2017}, and colloidal quantum dots \cite{Kagan2021}.

Solid-state SPE are usually identified and characterized using confocal microscopy, which facilitates the isolation of optical-diffraction-limited features from background fluorescence in a three-dimensional sample.
Typically, confocal microscopy images are analyzed by eye for features of interest.
This manual procedure is sufficient for the study of a few emitters, but efforts to screen engineered defects or bulk crystals to study the SPE's formation or statistical properties underscore the need for an automatic detection method.
Analyzing engineered defects created through ion implantation \cite{Wang2006, Toyli2010, Hollenbach2020, Rodt2020, Klein2020, Lagomarsino2021}, electron irradiation \cite{Fuchs2015, Capelli2019},  annealing \cite{Xu2018, Lyu2020}, or specialized growth schemes \cite{Palacios-Berraquero2017, Kim2019} often requires detecting hundreds of potential SPE. 
It is especially time-consuming to investigate randomly-located and often sparse emitters throughout three-dimensional (3D) bulk crystals, and thorough evaluation of all potential emitters is crucial for characterizing a novel material with unknown defect populations.

\begin{figure*} 
\includegraphics[trim={0 0 0 0}]{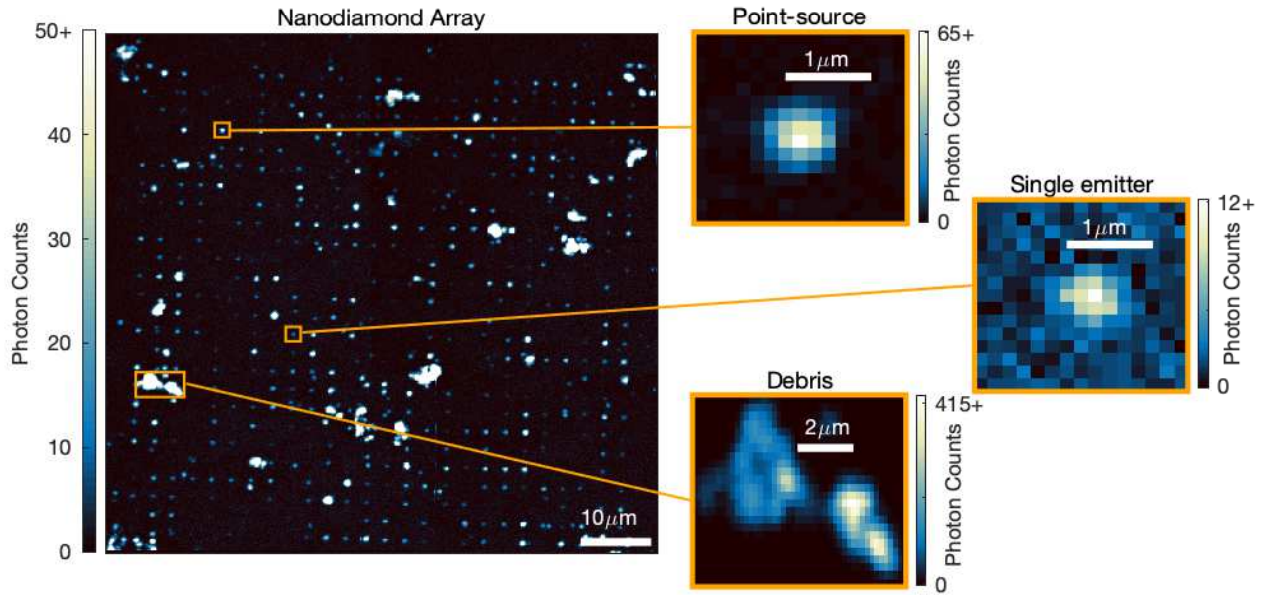}
    \caption{Stitched PL images of a nanodiamond array containing nitrogen-vacancy centers, acquired under 532 nm excitation. }
    \label{fig:1}
 \end{figure*}

Existing methods to characterize emitters in large-area PL images are primarily ensemble studies meant to assist manual detection. 
For instance, varying the excitation polarization of the PL image or applying a magnetic field highlights emitters with specific characteristics like potential spin states \cite{Exarhos2019}. 
Similar methods have been used to probe non-linear excitation by comparing two-photon and continuous wave PL images \cite{Schell2016} or by combining spectra and spatial intensity \cite{Stern2019}.
For materials with heterogeneous emitters, statistical modeling of the spatial distribution of emitter ensembles identifies defect families based on their intensity \cite{Breitweiser2020}.
However, these methods do not yield quantitative information regarding the optical properties of individual SPEs that can be used to classify or prioritize them for deeper investigation.. 
Once individual SPE of interest are identified, well-established methods exist to characterize their spatial, spectral, and temporal emission characteristics.
Examples of efficient characterization methods include wide-angle energy-momentum spectroscopy \cite{Dodson2014} and photon emission correlation spectroscopy \cite{fishman2021photon}.
Other methods to streamline the characterization of single emitters on an individual level include machine learning to reduce the time measuring the second-order autocorrelation function \cite{Kudyshev2020}. 
However, still missing from the literature is a flexible emitter screening step to evaluate individual emitters while also maintaining a broad view of emitter properties.

All defect-based SPE appear as point sources in confocal PL images, the shape of which is determined by the microscope's point spread function.
Ideally, the point spread function is a diffraction-limited Airy disk, but a 2D Gaussian function is typically an excellent approximation.
The Gaussian width depends on the numerical aperture of the confocal microscope setup and the emission wavelength. 
However, SPE are not identical across material systems.
Brightness and stability vary according to the characteristics of the defect and its host environment, with some emitters like those found in h-BN showing intermittent emission known as blinking \cite{VanDam2019, Chen2013, Cai2018a, Liu2015, GattoMonticone2014}. 
Promising emitters may bleach temporarily or permanently with continued laser exposure. 
Varying spatial distributions also create challenges for automatic detection. 
SPE can partially or completely overlap within the diffraction limit to yield a bright emitter that is indistinguishable by eye from true SPE \cite{Choi2016, Himics2014, Schroder2017, Thiruraman2019}. 
While overlap is naturally expected for spatially inhomogeneous emitter distributions created by focused ion beams or contained in dispersed nanoparticles,
emitters can cluster in apparently homogeneous samples due to invisible spatial variations in strain, composition, or extended structural defects.
Hence, an apparently isolated emitter is not guaranteed to be a single emitter.  
Finally, experimental factors such as image resolution, dwell time, and background signal intensity can all affect the ability to detect emitters. 
Thus, the heterogeneity of SPE systems warrants a highly flexible image processing method to accommodate the wide range of single emitter properties in PL images.

Here, we describe an automatic method that leverages the known optical properties of SPE to realize a widely-applicable screening criteria that emphasizes Gaussian shape, diffraction-limited size, and sufficient signal-to-noise ratio.  
The method accounts for the common features found in experimental PL images such as Figure~\ref{fig:1}, which includes point-sources, single-emitters, background signals, and debris. 
Our method analyzes the entire image in Figure~\ref{fig:1} in just 45 seconds, gathering quantitative information on each emitter's size, signal-to-noise, and shape to classify emitters according to their stability and intensity. 
The expected emitter characteristics, where known, inform which emitter group to prioritize for detailed studies, such as measurement of the second-order photon autocorrelation function to confirm single emission.  
Because the method groups similar emitters, the properties of confirmed SPE help to further refine the search.  
The result is a method that minimizes human input and reduces the time spent in the object detection and second-order autocorrelation stages of SPE exploration.
In addition to the array of nanodimonds containing nitrogen-vacancy (NV) centers \cite{shulevitz2021template} shown in Figure.~\ref{fig:1}, we apply the method to characterize SPE in exfoliated flakes of single-crystal h-BN.
The two samples are characterized by markedly different SPE properties, in terms of their spatial distribution, brightness, stability, and signal-to-noise ratio, and yet the method succeeds in rapidly classifying emitters in both cases.


\section{Methods}

\subsection{Experimental}
We use two custom-built confocal microscopes to image nanodiamond arrays containing NV centers \cite{shulevitz2021template} and exfoliated flakes of h-BN containing SPE.
The microscopes feature continuous-wave excitation lasers operating at 532 nm and 592 nm laser, respectively.
The nanodiamond PL image resolution is \SIrange{100}{150}{\nano\m} with 400 Hz scan rate, while the h-BN PL image resolution is \SI{50}{\nano\m} with 100 Hz scan rate. 
Emitters dominated by single-photon emission (hereafter called SPE) are confirmed by measuring the second-order photon autocorrelation function and determining a zero-delay value $g^{(2)}(\tau=0) < 0.5 $ without any background correction.
This metric may overestimate the number of single emitters \cite{fishman2021photon}, but it serves as a threshold for emitters worthy of more detailed investigation.  
Nanodiamonds are arranged in a grid with \SI{2.6}{\micro\m} spacing over Si/SiO$_2$ substrate.
\SI{25}{\micro\m} by \SI{25}{\micro\m} PL images of two nanodiamond arrays yield 192 emitters, 46 of which are deemed SPE.   
Bulk, single-crystal h-BN (acquired from HQ graphene) is mechanically exfoliated into thin ($<$\SI{100}{\nano\m}) flakes placed over Si/SiO$_2$ substrate fabricated with \SIrange{6}{8}{\micro\m} circular depressions recessed to a depth of \SI{5}{\micro\m}.
Emitters in h-BN are found around and within the recessed areas, and there are 10 verified SPE among two flakes. 

\subsection{Image Analysis}
The method we describe is related to established spectroscopic techniques such as single particle tracking \cite{Manzo2015} and super-resolution microscopy \cite{Schermelleh2019}, which fit emitters with a Gaussian function.
To start, 2D photoluminescence images are converted into binary images using an adaptive background threshold. 
The binary image is constructed by setting all pixels above or equal to the threshold to 1, while all pixels that are less than the threshold are set to 0. 
As opposed to a universal threshold, an adaptive threshold accounts for spatial variation in background intensity by calculating the local mean intensity for each pixel of the PL image \cite{Bradley2007}. 
Interconnected regions of foreground pixels that share at least an edge or corner are identified as individual objects.
Overlapping emitters within one interconnected region are separated by local maximum detection, and any objects smaller than the diffraction-limited spot-size are discarded. 
Regions centered on each object are cropped from the full PL image into at least 15 by 15 pixel regions of interest (ROI). 
Each detected object within the ROI is assigned a 2D Gaussian function within a simultaneous fit for all emitters plus a constant background.
The 2D symmetric Gaussian function is given by
\begin{equation} \label{eq:gauss}
    I(x,y) = A \cdot \exp\left(-\frac{(x-x_0)^2 + (y-y_0)^2}{2\sigma^2}\right),
\end{equation}
where $A$ is the peak intensity of the emitter, 
$(x_0,y_0)$ are the emitter coordinates, and $\sigma$ is the emitter width, corresponding to the Gaussian standard deviation.
The Gaussian fit provides the width, signal-to-noise ratio, and goodness-of-fit for each emitter, which serve as the primary detection criteria.

\section{Detection criteria}

\subsection{Width}
Objects are filtered based on their best-fit Gaussian width. 
For a diffraction-limited point-source emitter, the expected width is 
\begin{equation} \label{eq:width}
 \sigma_\mathrm{diff}=0.21\frac{\lambda}{\textrm{NA}},    
 \end{equation}
where $\lambda$ is the defect's emission wavelength and NA is the numerical aperture of the microscope objective \cite{Zhang2007}. 
If the emission wavelength is unknown, a width range can be estimated by substituting the microscope's detection range for $\lambda$. 
Typically, the detection range will be bound by the laser excitation wavelength and the the upper limit of the photon detector's range. 
In practice, optical aberrations or nonideal confocal conditions may distort the microscope's point spread function, increasing the apparent width.
This adjustment can often be calibrated using a multiplicative factor determined by comparison between the Gaussian function and the microscope's PSF, measured using a known sample with a bright, stable point source. 
We constrain the best-fit width to a range, $\sigma\in[\sigma_\mathrm{min},\sigma_\mathrm{max}]$, where $\sigma_\mathrm{min}$ and $\sigma_\mathrm{max}$ are set based on estimates for the expected Gaussian width corresponding to the wavelength range of interest and adjustments based on the microscope's point spread function.

\subsection{2D Gaussian Fit}
The symmetric 2D Gaussian fit filters the detected objects according to their shape, selecting for Gaussian point-sources over misshapen and extended objects. 
The  width constraint from the previous section is applied in this step as the range of allowed width for each peak in the fit.   
The emitter position is also tightly constrained to within 2 pixels of the weighted center of the detected object to ensure partially overlapping emitters are fit separately. 
We perform a least-squares regression fit where the free parameters include the position, width, and amplitude of each emitter plus a constant background.
The goodness-of-fit parameter is defined as reduced chi-squared ($\chi_R^2$), given by
\begin{equation} \label{eq:chi}
    \chi^2_R =\frac{1}{\mathrm{DoF}}\sum_{i=1}^{N} \frac{(O_i - M_i)^2}{\sigma_i^2}, 
\end{equation}
in which $M_i$ is the fitted counts, $O_i$ is the measured counts, $\sigma_i=\sqrt{O_i}$ is the Poisson noise of the measured counts, and $\mathrm{DoF}$ is the number of degrees of freedom in the fit. 

For an ideal fit with many degrees of freedom, $\chi_R^2$ equal to one means the model fits the data within the expected variance, $\chi_R^2$ less than one indicates the data is over-fitted by the model, and $\chi_R^2$ greater than one indicates a poor fit or a model that does not fully capture the data. 
The statistical expectation for a good fit depends on $\mathrm{DoF}$, with bounds at $ \chi_R^2 = 1\pm \sqrt{\frac{2}{\mathrm{DoF}}}$.
For ROIs containing 15$\times$15 pixels and one emitter, the expected $\chi_R^2 $ range based only on the degrees of freedom in the fit is 0.9 to 1.1. 
However, we have found that extending the allowed values on an empirical basis to 0.8 and 1.5 is necessary to account for other sources of error.
The upper limit accounts for slight under-fitting due to the systematic error between our microscope's PSF and the 2D Gaussian function. 
The lower limit reflects over-fitted emitters with low signal-to-noise ratio, which occurs when dim single emitters are close to the background level. 
A $\chi_R^2 $ value less than 0.8 indicates objects that are usually background fluctuations and should be discarded.
However, we find that a $\chi_R^2 $ value much greater than 1.5 is possible for bright single emitters affected by blinking and uneven PL image background. 
These emitters appear Gaussian and symmetric by eye except for a small number of dark pixels.
An uneven background or bright background object can also produce a large $\chi_R^2 $ value. 
Dim emitters are less affected by these factors due to their lower contrast with the background. 

Without an alternative measure, these bright, blinking emitters would be excluded on the basis of their high $\chi_R^2 $ value.
For materials with emitters frequently affected by these issues, we find that emitter shape\,---\,specifically, elliptical eccentricity\,---\,is an effective means to filter point sources from extended objects.
The elliptical 2D Gaussian function is defined as
\begin{equation}
      I(x,y) = A \cdot \exp\left(-(a  \tilde{x}^2 + 2b  \tilde{x}\tilde{y} + c  \tilde{y}^2)\right),
\end{equation}
where $(\tilde{x},\tilde{y})=(x-x_0,y-y_0)$ are relative coordinates, and
\begin{subequations}
\begin{align}
      a = & \frac{\cos^2\theta}{2  \sigma_1^2} + \frac{\sin^2\theta}{2 \sigma_2^2}, \\
    b = & \frac{-\sin^2 2\theta}{4 \sigma_1^2} + \frac{\sin^2 2\theta}{4  \sigma_2^2}, \\
    c = & \frac{\sin^2 \theta}{2  \sigma_1^2} + \frac{\cos^2 \theta }{2 \sigma_2^2},
\end{align}
\end{subequations}
where $\sigma_1$ and $\sigma_2$ are the Gaussian widths of the elliptical axes, and $\theta$ is the rotation angle.
The emitter eccentricity is defined as 
\begin{equation}
    e = \sqrt{1-\frac{\min[\sigma_1,\sigma_2]^2}{\max[\sigma_1,\sigma_2]^2}}.
\end{equation}
Possible eccentricity values range from 0 for a circle to 1 for a line.
We set the allowed range of eccentricity for SPE detection to any value between 0 and 0.66, corresponding to at least a 3:4 ratio between the minor and major widths. 
For bright emitters in materials known to have uneven background and frequent blinking, eccentricity overrides $\chi_R^2$ in classifying SPE. 

\begin{table*} 
    \centering
\begin{center}
    \begin{tabular}{ |c|c|c|c| }
        \hline
        \textbf{Group A} & \textbf{Group B} & \textbf{Group C} & \textbf{Group D} \\
        \hline
        $SNR>10$ & $2<SNR<10$ & $2<SNR<10$ & $SNR>10$ \\
        $0.8<\chi_R^2<1.5$ & $0.8<\chi_R^2<1.5$ & $0.8<\chi_R^2<1.5$ & $0 < $ e $ < 0.66$ \\
        Width within limits & Width within limits &  Width at limits & Width within limits \\
        \hline
        \includegraphics[scale =0.8]{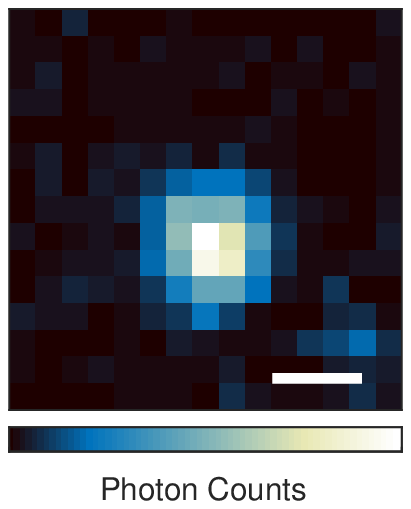} &
          \includegraphics[scale =0.8]{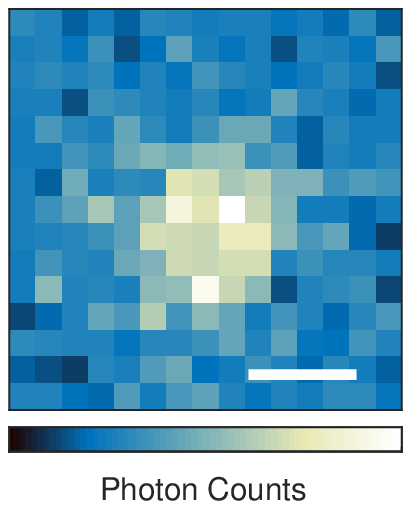} & \includegraphics[scale =0.8]{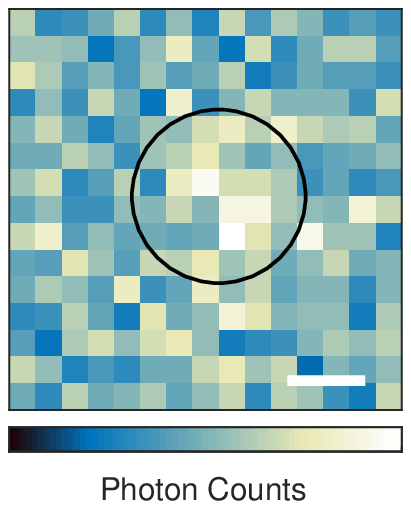} & 
          \includegraphics[scale =0.8]{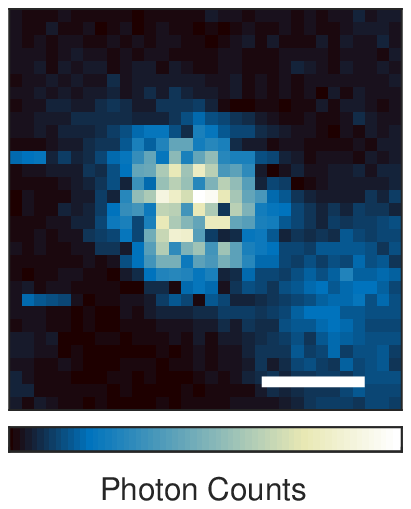} \\
          \hline
        bright, stable & dim, stable &  dim, blinking &  bright, blinking \\
          \hline
    \end{tabular}
       \caption{Emitter groups and representative single emitters observed in PL images of nanodiamond and h-BN. The examples shown for A and D are single emitters in h-BN, while those shown for C and B are single emitters from the nanodiamond array. PL counts are normalized to the maximum intensity of each image. Scale bars are \SI{0.5}{\micro\m}.}
       \label{tab:1}
       \end{center}
\end{table*}

\subsection{Signal-to-Noise Ratio}
The signal-to-noise ratio (SNR) is incorporated into the detection criteria to avoid the selection of dim, under-fitted emitters that are likely to be fluctuations in the background rather than single emitters. The SNR is given by
\begin{equation} \label{eq:snr}
\mathrm{SNR} =\frac{A}{\sqrt{B}} 
 \end{equation}
where $A$ is the best-fit emitter intensity and $B$ is the best-fit background value.
The SNR depends on both the emitter brightness and the acquisition settings (particularly the dwell time per pixel).
An alternative approach could use the signal-to-background ratio, which is less dependent on the acquisition settings, however we find that SNR is more robust in identifying SPE by incorporating both the emitter intensity and its statistical significance as an emitter defined above the background level. 
We set a lower limit of $\mathrm{SNR}>2$ for emitter detection, and a threshold of $\mathrm{SNR}=10$ to divide bright ($\mathrm{SNR}>10$) and dim ($\mathrm{SNR}<10$) emitters.

\section{Emitter Classification}
The detection criteria described in the previous section define four groups of emitters exhibiting qualitatively different characteristics. 
Table \ref{tab:1} lists the SNR, $\chi_R^2$, and width criteria corresponding to each emitter group, along with representative example images. 
All emitters classified in groups A-D are consistent with Gaussian point-sources. 
The groups are designed to capture similar emitters, reflecting common factors that affect PL images of single emitters like emitter instability and uneven background. 
Group A describes well-isolated, stable, bright objects, representing the ideal case for single emitter discovery. 
Group B emitters are identical to group A, except they are dim. 
Group C describes emitters with a good Gaussian fit but unexpectedly large or small width, which often occurs for dim and blinking emitters. 
Group D captures well-sized and shaped emitters for which background conditions or high frequency blinking significantly affect $\chi_R^2$.
By categorizing potential emitters in this manner, basic knowledge of the sample can direct further investigation into the most promising group. 

 \section{Results}
 \begin{figure}[b]
     \centering
     \includegraphics{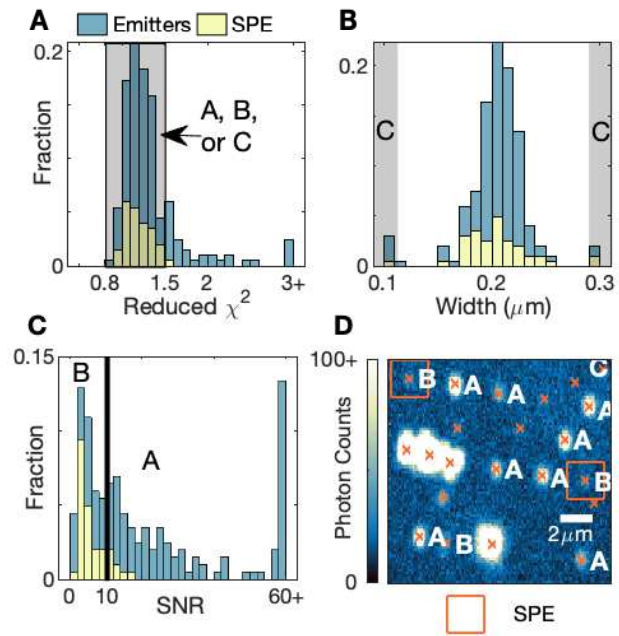}
     \caption{Emitter detection applied to 2D photoluminescence images of nanodiamond arrays. A) $\chi_R^2 $ distribution and B) Gaussian width distribution for SPE and non-SPE demonstrate how most non-SPE are Gaussian point-sources for this sample. C) SNR distribution showing intensity difference between SPE and non-SPE. D) Object detection and classification of emitters according to Table \ref{tab:1} for a portion of a nanodiamond array. Scale bar is 2 $\mu$m.}
     \label{fig:2}
 \end{figure}
 
The automatic detection method was developed and tested using nanodiamonds and h-BN to span the range of emitter and background properties for SPE in PL images. 
The nanodiamond array features regularly-spaced emitters placed at approximately the same depth. 
The NV centers contained in the nanodiamonds are stable, homogeneous, well-known SPE, and they provide a baseline for the ideal case of stable, well-isolated emitters over even background.
However stochastic distribution of NV centers within the nanodiamonds means that most spots contain multiple emitters.
In contrast, emitters in h-BN are more uniformly spatially distributed, whereas they are frequently unstable, heterogenous, and the flake itself may vary in thickness over the imaging area. 
The method accommodates both extremes.  

Figure~\ref{fig:2} depicts the application of this method to the nanodiamond array.
Figure~\ref{fig:2}A demonstrates the efficacy of the chosen $\chi_R^2$ range in identifying SPE, and it further highlights how many objects in this sample are similarly well-shaped point-sources but not single emitters.
Figure~\ref{fig:2}B demonstrates how almost all emitters in the nanodiamond array fall well within the chosen width limits, set as \SIrange{100}{300}{\nano\m} to accommodate noise and systematic errors in the confocal setup. 
A few single emitters reach the limits and fall into group C (dim, unstable).  
Figure~\ref{fig:2}B also clarifies the extent of systematic error in our confocal setup.
The approximate Gaussian width for a single NV center should be \SI{152}{\nano\m}, calculated using Eq.~(\ref{eq:width}) for \SI{690}{\nano\m} peak emission with \SI{532}{\nano\m} excitation \cite{Fukushige2020} and NA equal to $0.95$. 
However, the average Gaussian width in figure~\ref{fig:2}B for NV centers in our confocal setup is \SI{203}{\nano\m}, with $\sim$95$\%$ of emitters falling within $\pm$\SI{60}{\nano\m} of the mean. 
This error reflects the discrepancy between our microscope's PSF and the ideal diffraction limit and motivates the extended width range. 
The final step, shown in Figure~\ref{fig:2}C, is to differentiate stable emitters based on intensity into groups A (bright, stable) and B (dim, stable). 
Due to the stability of the NV centers in this sample and the low, uniform background of the silicon substrate, these images do not benefit from the elliptical Gaussian fit, which is designed to gauge symmetry for blinking emitters or emitters over uneven background. 
Therefore, group D (bright, blinking) is not required. 

\begin{figure}[t]
     \centering
     \includegraphics{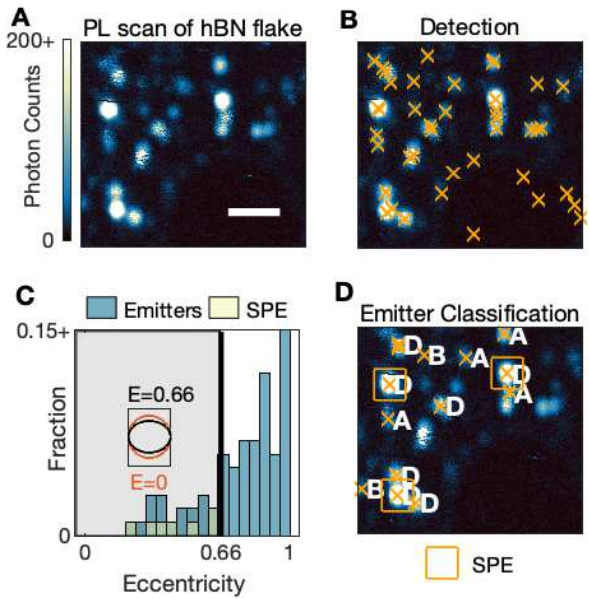}
     \caption{Emitter detection for hexagonal boron nitride flake. Scale bar is \SI{2.5}{\micro\m}. A) PL image acquired under \SI{592}{\nano\m} excitation at \SI{245}{\micro\W}. B) Object detection with adaptive thresholding. C) Distribution of eccentricity according to the elliptical Gaussian fit for SPE and other emitters found in two h-BN flakes, including the flake pictured in A. D) Emitter classification of the objects detected in B according to groups outlined in Table \ref{tab:1}. }
     \label{fig:3}
 \end{figure}
 
Figure~\ref{fig:3} displays the image analysis and emitter classification for an h-BN flake imaged with a \SI{592}{\nano\m} laser. 
The PL scan in Figure~\ref{fig:3}A shows overlapping and blinking emitters with an uneven background.
In Figure~\ref{fig:3}B, the object detection step identifies possible emitters despite these nonidealities.
The prevalence of blinking emitters in this sample motivates the use of shape analysis to capture symmetric but unstable emitters.
In Figure~~\ref{fig:3}C, we show the eccentricity distribution for the emitters detected in this sample along with the threshold, $e<0.66$.
Notably, the majority of detected emitters fall outside this range and are excluded on the basis of likely being emitter clusters or extended objects.
Figure~\ref{fig:3}D labels each emitter retained by the analysis with its classification type; the image shows a dominance of group D, demonstrating the utility of including eccentricity as a secondary shape measure for capturing unstable emitters that would otherwise be excluded.

The fluorescent defects in h-BN are generally heterogeneous in structure and feature a range of emission wavelengths \cite{Exarhos2017}. 
Thus, the emission wavelength cannot be estimated consistently to give the expected size of a diffraction-limited spot. 
Width limits are set instead according to the detection range of the setup, determined by the \SI{645}{\nano\m} long wave pass (LWP) filter in the setup and the decline of the photon detector efficiency at $\sim$\SI{1000}{\nano\m}.
This range corresponds to \SIrange{150}{233}{\nano\m} emitter width using Eq.~(\ref{eq:width}), and the width limits are set from \SIrange{100}{350}{\nano\m} to reflect the empirically broadened distribution revealed in the nanodiamond data (Figure~\ref{fig:2}B).
 
\begin{figure}[b]
    \centering
    \includegraphics{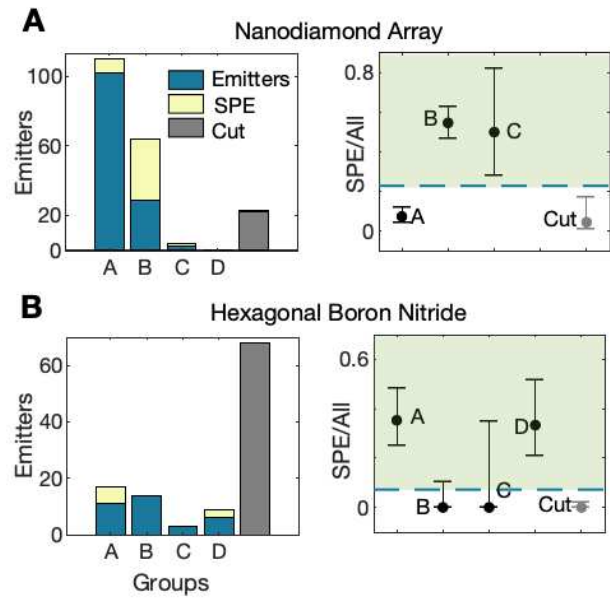}
    \caption{SPE distribution for group A (bright and stable), group B (dim and stable), group C (dim and unstable), and group D (bright and unstable) for A) the nanodiamond arrays and B) the h-BN samples. Blue dashed line denotes the total fraction of SPE for all emitters (0.227 for nanodiamonds and 0.072 for h-BN). Error bars indicate the Wilson confidence interval. }
    \label{fig:4}
\end{figure}
 
Figure~\ref{fig:4} summarizes the performance of the method for the nanodiamond and h-BN data. 
For each sample, we plot the distribution of emitters grouped in each classification (including those that were excluded), and indicate the number of SPE.
The emitter groups substantially improve the probability of finding SPE compared to random sampling of all potential emitters. 
Furthermore, the emitters discarded or cut by the method show a very low likelihood of being SPE. 
The uncertainty in the predicted concentration of SPE within each group is given by the Wilson interval \cite{Wilson1927} to reflect the statistical uncertainty based on the number of emitters within each group. 

For the nanodiamond arrays, 22\% of all emitters are SPE, but group B contains 54\% SPE.
Indeed, SPE are expected to be relatively dim and stable (group B) in this sample since most nanodiamonds contain more than one NV center.
For the h-BN sample, only 7$\%$ of all potential emitters are SPE, but groups A and D each contain over 30$\%$ SPE. 
Notably, many bright and stable emitters (group A) in h-BN are not single emitters, suggesting a higher degree of emitter clustering than would be expected for randomly distributed defects. 
Group C is least common for both the nanodiamonds and h-BN samples, which is reflected in the large uncertainty range in Figure ~\ref{fig:4}.

\section{Discussion} \label{section:discussion} 
\subsection{Accounting for Sample, Emitter, and Setup Variation}
The detection criteria we describe are rooted in the fundamental optical properties of fluorescent defects, but in practice there are several systematic and random sources of error that cause the measured PL of the emitter to stray from the theoretical ideal.
It is necessary to understand and anticipate error to tailor the method to unique materials and setups.

Systematic error sources include the microscope's point-spread function, position-dependent optical aberrations, and material-specific properties like high or uneven background and blinking emitters.  
Random error originates from the Poisson noise of the measured counts and is exacerbated by low PL image acquisition time and resolution. 
The method accounts for these errors by expanding the limits of the detection criteria and tolerating false positives to ensure that very few SPE are missed.
The specific error implications for the width, Gaussian fit, and SNR criteria are discussed in this section. 

\subsubsection{Width}
The width criterion is strongly affected by setup-dependent factors that create a discrepancy between the fitted Gaussian width and the expected size of a diffraction-limited spot. 
In the nanodiamond sample, for example, the expected spot size in our confocal setup is between \SIrange{117}{233}{\nano\m}, as calculated from Eq.~(\ref{eq:width}), with $\lambda$ bounded by the laser excitation (\SI{532}{\nano\m}) and the upper limit of the photon detector (\SI{1000}{\nano\m}). 
However, confirmed SPE in the nanodiamond arrays display widths between \SIrange{100}{300}{\nano\m} (Figure~\ref{fig:2}B). 
Systematic broadening can be quantified by comparing a known point source like the NV center with the Gaussian PSF, providing a scaling factor for the expected width.
However, without time-consuming, high-resolution PL images, the large width variance due to random error still prevents a strict width limit. 
In practice, it is also likely that emission wavelength of each emitter will initially be unknown.  
In this case, the most effective strategy is to screen for emitters that are reasonably within the diffraction limit given the error of the setup. 
The broad width limits applied for the nanodiamond and h-BN samples ensured that no SPE were missed and successfully filtered small background fluctuations and large, extended objects. 

Any emitters that comply with the SNR and $\chi_R^2$ criteria but reach the imposed width limits require special consideration. 
These emitters are typically dim and blinking, presenting as just a few bright pixels or a partial 2D Gaussian PSF. 
Blinking SPE may still be well-fit by the 2D Gaussian PSF if the emitter is close to the background level, but the width may not accurately reflect the emitter. 
Thus, low SNR emitters with unusual widths are isolated into group C to prioritize emitter stability and acknowledge that in rare cases, the width-limited fit may not fully describe the emitter.
A small number of SPE in the nanodiamond arrays fall into group C. 

\subsubsection{Goodness of Fit}
The inherent discrepancy between the microscope's PSF and the 2D Gaussian fitting function will shift the expected $\chi_R^2$, even for a perfect point-source emitter. 
In our confocal setup, we account for this discrepancy by moving the upper limit of $\chi_R^2$ to 1.5, which is higher than the limit predicted on purely statistical grounds. 
This limit of 1.5 is likely applicable to other setups with similar acquisition settings unless there is a high degree of shape distortion visible by eye. 
The precise upper limit may be adjusted once a sample of known SPE are acquired and fit. 
Similarly, while the 0 to 0.66 eccentricity range used to define group D reflects the properties of our confocal microscope, this range will likely be reasonable for another well-aligned setup.
It is also possible to reduce the allowed range of eccentricity to define a more selective group D. 

The selectivity of the Gaussian fit is strongly impacted by material characteristics such as poor background, low signal-to-noise (2$<$SNR$<$5), and blinking. 
Blinking points and uneven background bias the Gaussian fit because each point's contribution to $\chi_R^2$ is weighted by the inverse of the variance of the measured photon counts. 
Therefore, a poor fit among dim background or blinking points disproportionately worsens $\chi_R^2$, with blinking SPE in h-BN sometimes showing $\chi_R^2$ up to 30.  
Group D captures bright and symmetric emitters that would be otherwise missed by the method because of background or blinking effects.   

\subsubsection{Signal-to-Noise Ratio}
The SNR criterion is least affected by both setup and material sources of error, however they do depend somewhat on the acquisition settings.  
The SNR computation relies on the Gaussian fit for the peak intensity of the emitter and background level, but these parameters are accurate if the fit is constrained to the emitter and the initial background condition is reasonable.
In our images, the definitions of bright (SNR$>$10) and dim (2$<$SNR$<$10) successfully separate most point-source emitters from single emitters in the nanodiamond sample as shown in Figure~\ref{fig:2}C.
These limits also distinguish bright, blinking emitters in h-BN (Figure~\ref{fig:3}) for classification as group D.
Similar limits are likely applicable to any material system, as they describe the emitter's relation to the background as it pertains to the detection criteria. 
For example, any bright emitter over locally uneven background will benefit from the width ratio calculation to double-check the circularity.  
If needed, the precise thresholds can be adjusted to reflect the overall confocal image SNR as determined by its resolution and dwell time.

\subsection{Practical Use Considerations}
When applying the method to a particular sample, the objective is to determine which group will be the best to prioritize for SPE search.
Systems with stable, isolated, and high intensity emitters, including color centers in diamond and quantum dots, will tend to exhibit SPE in the stable and bright group A.
For ion-implanted samples or fabricated nanoparticle arrays, defects are naturally clustered within diffraction-limited spots. 
Thus, single emitters are likely to be found in the dim C and B groups.
Samples of this type include the nanodiamond arrays, etched arrays of nanopillars  \cite{McCloskey2020} and nanopyramids \cite{Jaffe2019}, and ion-implanted arrays.
A high occurrence of emitter blinking should guides user to implement the group D categorization based on shape, especially if there are very few emitters categorized as the bright and stable group A. 
H-BN is a prominent example of blinking emitters, but other examples include nanocrystals \cite{Bradac2010, Nirmal1996} and certain quantum dots \cite{Davanco2014}. 

For lesser-known or entirely new samples, it is more difficult to predict the expected characteristics of SPE.
However, basic knowledge of the sample can provide some guidance to prioritize autocorrelation measurements for certain groups. 
First, the user should determine the applicability of group D, as the elliptical Gaussian fit approximately doubles the time to apply the method.  
Group D should be used only if initial large-area PL images show a high frequency of blinking, or if the method discards most emitters with only groups A-C. 
Even if group D is engaged, group A emitters should probably take precedence for measurements because their stability and high signal-to-noise will enable shorter acquisition times. 
Group D emitters could be investigated next, since these emitters also exhibit high signal-to-noise ratios.  
However, if the user expects emitter overlap within the diffraction limit, the dim and stable group B should be prioritized for potential SPE instead of groups A or D. 
Group C should likely be investigated last, as dim and blinking emitters do not generally possess favorable properties for SPE. 
For samples with no expectation of overlapping emitters, the dim groups B and C will contain emitters that are close to the background and require long autocorrelation acquisition times.

This method is designed to identify diffraction-limited point-sources, and thus a high prevalence of false positives is largely dependent on emitter spatial clustering. 
For instance, the high rate of nanodiamond false positives shown in Figure~\ref{fig:4} is expected because the NV centers are naturally confined within sub-diffraction ($\sim$\SI{40}{nm}-diameter) particles.
Acquisition settings are also important in the method's efficacy, and a certain amount of trial-and-error can help to optimize the performance.
Low PL image resolution can lead false positives that are only background fluctuations. 
In general, the PL image resolution should be no larger than the mean of the expected width range.
Resolution at half the mean of the expected width or below is recommended for samples with unstable, crowded emitters over inconsistent background.
Photo-bleaching the sample can also help reduce the the background and the occurrence of false positives. 

\section{Conclusion}
The method presented in this paper provides a general, flexible framework for efficiently screening new materials for SPE. 
Compared to manual searching, the ability to detect and classify individual fluorescent emitters through quick image processing greatly improves the speed of exploratory experiments.
While we have demonstrated the utility of this method for commonly studied defect hosts like h-BN and diamond, it can also be applied to promising yet largely unexplored materials. 
Potential new hosts include compound semiconductors like magnesium and strontium oxides, group II-VI materials like zinc sulfide, complex metal oxides like yttrium orthosilicate, and perovskites \cite{Bassett2019}.  
While we consider two-dimensional confocal images of planar samples, the method can be readily adapted to detect emitters distributed within three-dimensional samples.
This can be accomplished by automating acquisition of PL maps through different planes and performing sequential or simultaneous analysis of each image.
In addition to investigating new SPE, the emitter detection and characterization method can also generally enhance knowledge of the defect and material system,
by providing quantitative information about the emitter spatial locations, brightness, and shape.
For instance, the correlation of defects' optical properties and positions can establish relationships between the existence and properties of SPE with grain boundaries, growth sectors, or other material features.
The process to optimize ion-implantation dose and energy for single defect creation is also improved by efficient detection and characterization of large emitter arrays. 
The method could also be adapted to classify emitters based on additional spectral, temporal, or polarization-dependent information that is acquired through more advanced imaging techniques.
In summary, this method for rapid characterization of emitters in PL images will advance the discovery and study of new defects for quantum technology applications, and it can help to expand the collective knowledge of quantum defects and the material factors that influence their properties.

\section{Acknowledgements}
This work was supported by the National Science Foundation under Award DMR-1922278.
We thank Tzu-Yung Huang for helpful discussions about optical aberrations and image processing.
L.R.N. acknowledges support from the University Scholars Program at the University of Pennsylvania.
H.J.S. and L.C.B. acknowledge support from NSF through the University of Pennsylvania Materials Research Science and Engineering Center (MRSEC) (DMR-1720530) to develop the nanodiamond arrays used in this study.

\bibliography{library.bib}

\begin{thebibliography}{70}%
\makeatletter
\providecommand \@ifxundefined [1]{%
 \@ifx{#1\undefined}
}%
\providecommand \@ifnum [1]{%
 \ifnum #1\expandafter \@firstoftwo
 \else \expandafter \@secondoftwo
 \fi
}%
\providecommand \@ifx [1]{%
 \ifx #1\expandafter \@firstoftwo
 \else \expandafter \@secondoftwo
 \fi
}%
\providecommand \natexlab [1]{#1}%
\providecommand \enquote  [1]{``#1''}%
\providecommand \bibnamefont  [1]{#1}%
\providecommand \bibfnamefont [1]{#1}%
\providecommand \citenamefont [1]{#1}%
\providecommand \href@noop [0]{\@secondoftwo}%
\providecommand \href [0]{\begingroup \@sanitize@url \@href}%
\providecommand \@href[1]{\@@startlink{#1}\@@href}%
\providecommand \@@href[1]{\endgroup#1\@@endlink}%
\providecommand \@sanitize@url [0]{\catcode `\\12\catcode `\$12\catcode
  `\&12\catcode `\#12\catcode `\^12\catcode `\_12\catcode `\%12\relax}%
\providecommand \@@startlink[1]{}%
\providecommand \@@endlink[0]{}%
\providecommand \url  [0]{\begingroup\@sanitize@url \@url }%
\providecommand \@url [1]{\endgroup\@href {#1}{\urlprefix }}%
\providecommand \urlprefix  [0]{URL }%
\providecommand \Eprint [0]{\href }%
\providecommand \doibase [0]{https://doi.org/}%
\providecommand \selectlanguage [0]{\@gobble}%
\providecommand \bibinfo  [0]{\@secondoftwo}%
\providecommand \bibfield  [0]{\@secondoftwo}%
\providecommand \translation [1]{[#1]}%
\providecommand \BibitemOpen [0]{}%
\providecommand \bibitemStop [0]{}%
\providecommand \bibitemNoStop [0]{.\EOS\space}%
\providecommand \EOS [0]{\spacefactor3000\relax}%
\providecommand \BibitemShut  [1]{\csname bibitem#1\endcsname}%
\let\auto@bib@innerbib\@empty
\bibitem [{\citenamefont {Childress}\ and\ \citenamefont
  {Hanson}(2013)}]{Childress2013}%
  \BibitemOpen
  \bibfield  {author} {\bibinfo {author} {\bibfnamefont {L.}~\bibnamefont
  {Childress}}\ and\ \bibinfo {author} {\bibfnamefont {R.}~\bibnamefont
  {Hanson}},\ }\bibfield  {title} {\bibinfo {title} {{Diamond NV centers for
  quantum computing and quantum networks}},\ }\href
  {https://doi.org/10.1557/mrs.2013.20} {\bibfield  {journal} {\bibinfo
  {journal} {MRS Bulletin}\ }\textbf {\bibinfo {volume} {38}},\ \bibinfo
  {pages} {134} (\bibinfo {year} {2013})}\BibitemShut {NoStop}%
\bibitem [{\citenamefont {Liu}\ and\ \citenamefont {Hersam}(2019)}]{Liu2019}%
  \BibitemOpen
  \bibfield  {author} {\bibinfo {author} {\bibfnamefont {X.}~\bibnamefont
  {Liu}}\ and\ \bibinfo {author} {\bibfnamefont {M.~C.}\ \bibnamefont
  {Hersam}},\ }\bibfield  {title} {\bibinfo {title} {{2D materials for quantum
  information science}},\ }\href {https://doi.org/10.1038/s41578-019-0136-x}
  {\bibfield  {journal} {\bibinfo  {journal} {Nature Reviews Materials}\
  }\textbf {\bibinfo {volume} {4}},\ \bibinfo {pages} {669} (\bibinfo {year}
  {2019})}\BibitemShut {NoStop}%
\bibitem [{\citenamefont {Jiang}\ \emph {et~al.}(2007)\citenamefont {Jiang},
  \citenamefont {Taylor}, \citenamefont {S{\o}rensen},\ and\ \citenamefont
  {Lukin}}]{Jiang2007}%
  \BibitemOpen
  \bibfield  {author} {\bibinfo {author} {\bibfnamefont {L.}~\bibnamefont
  {Jiang}}, \bibinfo {author} {\bibfnamefont {J.~M.}\ \bibnamefont {Taylor}},
  \bibinfo {author} {\bibfnamefont {A.~S.}\ \bibnamefont {S{\o}rensen}},\ and\
  \bibinfo {author} {\bibfnamefont {M.~D.}\ \bibnamefont {Lukin}},\ }\bibfield
  {title} {\bibinfo {title} {{Distributed quantum computation based on small
  quantum registers}},\ }\href {https://doi.org/10.1103/PhysRevA.76.062323}
  {\bibfield  {journal} {\bibinfo  {journal} {Phys. Rev. A}\ }\textbf {\bibinfo
  {volume} {76}},\ \bibinfo {pages} {062323} (\bibinfo {year}
  {2007})}\BibitemShut {NoStop}%
\bibitem [{\citenamefont {Atat{\"{u}}re}\ \emph {et~al.}(2018)\citenamefont
  {Atat{\"{u}}re}, \citenamefont {Englund}, \citenamefont {Vamivakas},
  \citenamefont {Lee},\ and\ \citenamefont {Wrachtrup}}]{Atature2018}%
  \BibitemOpen
  \bibfield  {author} {\bibinfo {author} {\bibfnamefont {M.}~\bibnamefont
  {Atat{\"{u}}re}}, \bibinfo {author} {\bibfnamefont {D.}~\bibnamefont
  {Englund}}, \bibinfo {author} {\bibfnamefont {N.}~\bibnamefont {Vamivakas}},
  \bibinfo {author} {\bibfnamefont {S.~Y.}\ \bibnamefont {Lee}},\ and\ \bibinfo
  {author} {\bibfnamefont {J.}~\bibnamefont {Wrachtrup}},\ }\bibfield  {title}
  {\bibinfo {title} {{Material platforms for spin-based photonic quantum
  technologies}},\ }\href {https://doi.org/10.1038/s41578-018-0008-9}
  {\bibfield  {journal} {\bibinfo  {journal} {Nature Reviews Materials}\
  }\textbf {\bibinfo {volume} {3}},\ \bibinfo {pages} {38} (\bibinfo {year}
  {2018})}\BibitemShut {NoStop}%
\bibitem [{\citenamefont {Degen}\ \emph {et~al.}(2017)\citenamefont {Degen},
  \citenamefont {Reinhard},\ and\ \citenamefont {Cappellaro}}]{Degen2017}%
  \BibitemOpen
  \bibfield  {author} {\bibinfo {author} {\bibfnamefont {C.~L.}\ \bibnamefont
  {Degen}}, \bibinfo {author} {\bibfnamefont {F.}~\bibnamefont {Reinhard}},\
  and\ \bibinfo {author} {\bibfnamefont {P.}~\bibnamefont {Cappellaro}},\
  }\bibfield  {title} {\bibinfo {title} {{Quantum sensing}},\ }\href
  {https://doi.org/10.1103/RevModPhys.89.035002} {\bibfield  {journal}
  {\bibinfo  {journal} {Reviews of Modern Physics}\ }\textbf {\bibinfo {volume}
  {89}},\ \bibinfo {pages} {035005} (\bibinfo {year} {2017})}\BibitemShut
  {NoStop}%
\bibitem [{\citenamefont {Schirhagl}\ \emph {et~al.}(2014)\citenamefont
  {Schirhagl}, \citenamefont {Chang}, \citenamefont {Loretz},\ and\
  \citenamefont {Degen}}]{Schirhagl2014}%
  \BibitemOpen
  \bibfield  {author} {\bibinfo {author} {\bibfnamefont {R.}~\bibnamefont
  {Schirhagl}}, \bibinfo {author} {\bibfnamefont {K.}~\bibnamefont {Chang}},
  \bibinfo {author} {\bibfnamefont {M.}~\bibnamefont {Loretz}},\ and\ \bibinfo
  {author} {\bibfnamefont {C.~L.}\ \bibnamefont {Degen}},\ }\bibfield  {title}
  {\bibinfo {title} {{Nitrogen-vacancy centers in diamond: Nanoscale sensors
  for physics and biology}},\ }\href
  {https://doi.org/10.1146/annurev-physchem-040513-103659} {\bibfield
  {journal} {\bibinfo  {journal} {Annual Review of Physical Chemistry}\
  }\textbf {\bibinfo {volume} {65}},\ \bibinfo {pages} {83} (\bibinfo {year}
  {2014})}\BibitemShut {NoStop}%
\bibitem [{\citenamefont {Heshami}\ \emph {et~al.}(2016)\citenamefont
  {Heshami}, \citenamefont {England}, \citenamefont {Humphreys}, \citenamefont
  {Bustard}, \citenamefont {Acosta}, \citenamefont {Nunn},\ and\ \citenamefont
  {Sussman}}]{Heshami2016}%
  \BibitemOpen
  \bibfield  {author} {\bibinfo {author} {\bibfnamefont {K.}~\bibnamefont
  {Heshami}}, \bibinfo {author} {\bibfnamefont {D.~G.}\ \bibnamefont
  {England}}, \bibinfo {author} {\bibfnamefont {P.~C.}\ \bibnamefont
  {Humphreys}}, \bibinfo {author} {\bibfnamefont {P.~J.}\ \bibnamefont
  {Bustard}}, \bibinfo {author} {\bibfnamefont {V.~M.}\ \bibnamefont {Acosta}},
  \bibinfo {author} {\bibfnamefont {J.}~\bibnamefont {Nunn}},\ and\ \bibinfo
  {author} {\bibfnamefont {B.~J.}\ \bibnamefont {Sussman}},\ }\bibfield
  {title} {\bibinfo {title} {{Quantum memories: emerging applications and
  recent advances}},\ }\href {https://doi.org/10.1080/09500340.2016.1148212}
  {\bibfield  {journal} {\bibinfo  {journal} {Journal of Modern Optics}\
  }\textbf {\bibinfo {volume} {63}},\ \bibinfo {pages} {2005} (\bibinfo {year}
  {2016})}\BibitemShut {NoStop}%
\bibitem [{\citenamefont {Bradley}\ \emph {et~al.}(2019)\citenamefont
  {Bradley}, \citenamefont {Randall}, \citenamefont {Abobeih}, \citenamefont
  {Berrevoets}, \citenamefont {Degen}, \citenamefont {Bakker}, \citenamefont
  {Markham}, \citenamefont {Twitchen},\ and\ \citenamefont
  {Taminiau}}]{Bradley2019}%
  \BibitemOpen
  \bibfield  {author} {\bibinfo {author} {\bibfnamefont {C.~E.}\ \bibnamefont
  {Bradley}}, \bibinfo {author} {\bibfnamefont {J.}~\bibnamefont {Randall}},
  \bibinfo {author} {\bibfnamefont {M.~H.}\ \bibnamefont {Abobeih}}, \bibinfo
  {author} {\bibfnamefont {R.~C.}\ \bibnamefont {Berrevoets}}, \bibinfo
  {author} {\bibfnamefont {M.~J.}\ \bibnamefont {Degen}}, \bibinfo {author}
  {\bibfnamefont {M.~A.}\ \bibnamefont {Bakker}}, \bibinfo {author}
  {\bibfnamefont {M.}~\bibnamefont {Markham}}, \bibinfo {author} {\bibfnamefont
  {D.~J.}\ \bibnamefont {Twitchen}},\ and\ \bibinfo {author} {\bibfnamefont
  {T.~H.}\ \bibnamefont {Taminiau}},\ }\bibfield  {title} {\bibinfo {title} {{A
  Ten-Qubit Solid-State Spin Register with Quantum Memory up to One Minute}},\
  }\href {https://doi.org/10.1103/PhysRevX.9.031045} {\bibfield  {journal}
  {\bibinfo  {journal} {Physical Review X}\ }\textbf {\bibinfo {volume} {9}},\
  \bibinfo {pages} {031045} (\bibinfo {year} {2019})}\BibitemShut {NoStop}%
\bibitem [{\citenamefont {Gao}\ \emph {et~al.}(2015)\citenamefont {Gao},
  \citenamefont {Imamoglu}, \citenamefont {Bernien},\ and\ \citenamefont
  {Hanson}}]{Gao2015}%
  \BibitemOpen
  \bibfield  {author} {\bibinfo {author} {\bibfnamefont {W.~B.}\ \bibnamefont
  {Gao}}, \bibinfo {author} {\bibfnamefont {A.}~\bibnamefont {Imamoglu}},
  \bibinfo {author} {\bibfnamefont {H.}~\bibnamefont {Bernien}},\ and\ \bibinfo
  {author} {\bibfnamefont {R.}~\bibnamefont {Hanson}},\ }\bibfield  {title}
  {\bibinfo {title} {{Coherent manipulation, measurement and entanglement of
  individual solid-state spins using optical fields}},\ }\href
  {https://doi.org/10.1038/nphoton.2015.58} {\bibfield  {journal} {\bibinfo
  {journal} {Nature Photonics}\ }\textbf {\bibinfo {volume} {9}},\ \bibinfo
  {pages} {363} (\bibinfo {year} {2015})}\BibitemShut {NoStop}%
\bibitem [{\citenamefont {Wehner}\ \emph {et~al.}(2018)\citenamefont {Wehner},
  \citenamefont {Elkouss},\ and\ \citenamefont {Hanson}}]{Wehner2018}%
  \BibitemOpen
  \bibfield  {author} {\bibinfo {author} {\bibfnamefont {S.}~\bibnamefont
  {Wehner}}, \bibinfo {author} {\bibfnamefont {D.}~\bibnamefont {Elkouss}},\
  and\ \bibinfo {author} {\bibfnamefont {R.}~\bibnamefont {Hanson}},\
  }\bibfield  {title} {\bibinfo {title} {{Quantum internet: A vision for the
  road ahead}},\ }\href {https://doi.org/10.1126/science.aam9288} {\bibfield
  {journal} {\bibinfo  {journal} {Science}\ }\textbf {\bibinfo {volume}
  {362}},\ \bibinfo {pages} {303} (\bibinfo {year} {2018})}\BibitemShut
  {NoStop}%
\bibitem [{\citenamefont {Arakawa}\ and\ \citenamefont
  {Holmes}(2020)}]{Arakawa2020}%
  \BibitemOpen
  \bibfield  {author} {\bibinfo {author} {\bibfnamefont {Y.}~\bibnamefont
  {Arakawa}}\ and\ \bibinfo {author} {\bibfnamefont {M.~J.}\ \bibnamefont
  {Holmes}},\ }\bibfield  {title} {\bibinfo {title} {{Progress in quantum-dot
  single photon sources for quantum information technologies: A broad spectrum
  overview}},\ }\href {https://doi.org/10.1063/5.0010193} {\bibfield  {journal}
  {\bibinfo  {journal} {Applied Physics Reviews}\ }\textbf {\bibinfo {volume}
  {7}},\ \bibinfo {pages} {021309} (\bibinfo {year} {2020})}\BibitemShut
  {NoStop}%
\bibitem [{\citenamefont {Somaschi}\ \emph {et~al.}(2016)\citenamefont
  {Somaschi}, \citenamefont {Giesz}, \citenamefont {{De Santis}}, \citenamefont
  {Loredo}, \citenamefont {Almeida}, \citenamefont {Hornecker}, \citenamefont
  {Portalupi}, \citenamefont {Grange}, \citenamefont {Ant{\'{o}}n},
  \citenamefont {Demory}, \citenamefont {G{\'{o}}mez}, \citenamefont {Sagnes},
  \citenamefont {Lanzillotti-Kimura}, \citenamefont {Lema{\'{i}}tre},
  \citenamefont {Auffeves}, \citenamefont {White}, \citenamefont {Lanco},\ and\
  \citenamefont {Senellart}}]{Somaschi2016}%
  \BibitemOpen
  \bibfield  {author} {\bibinfo {author} {\bibfnamefont {N.}~\bibnamefont
  {Somaschi}}, \bibinfo {author} {\bibfnamefont {V.}~\bibnamefont {Giesz}},
  \bibinfo {author} {\bibfnamefont {L.}~\bibnamefont {{De Santis}}}, \bibinfo
  {author} {\bibfnamefont {J.~C.}\ \bibnamefont {Loredo}}, \bibinfo {author}
  {\bibfnamefont {M.~P.}\ \bibnamefont {Almeida}}, \bibinfo {author}
  {\bibfnamefont {G.}~\bibnamefont {Hornecker}}, \bibinfo {author}
  {\bibfnamefont {S.~L.}\ \bibnamefont {Portalupi}}, \bibinfo {author}
  {\bibfnamefont {T.}~\bibnamefont {Grange}}, \bibinfo {author} {\bibfnamefont
  {C.}~\bibnamefont {Ant{\'{o}}n}}, \bibinfo {author} {\bibfnamefont
  {J.}~\bibnamefont {Demory}}, \bibinfo {author} {\bibfnamefont
  {C.}~\bibnamefont {G{\'{o}}mez}}, \bibinfo {author} {\bibfnamefont
  {I.}~\bibnamefont {Sagnes}}, \bibinfo {author} {\bibfnamefont {N.~D.}\
  \bibnamefont {Lanzillotti-Kimura}}, \bibinfo {author} {\bibfnamefont
  {A.}~\bibnamefont {Lema{\'{i}}tre}}, \bibinfo {author} {\bibfnamefont
  {A.}~\bibnamefont {Auffeves}}, \bibinfo {author} {\bibfnamefont {A.~G.}\
  \bibnamefont {White}}, \bibinfo {author} {\bibfnamefont {L.}~\bibnamefont
  {Lanco}},\ and\ \bibinfo {author} {\bibfnamefont {P.}~\bibnamefont
  {Senellart}},\ }\bibfield  {title} {\bibinfo {title} {{Near-optimal
  single-photon sources in the solid state}},\ }\href
  {https://doi.org/10.1038/nphoton.2016.23} {\bibfield  {journal} {\bibinfo
  {journal} {Nature Photonics}\ }\textbf {\bibinfo {volume} {10}},\ \bibinfo
  {pages} {340} (\bibinfo {year} {2016})}\BibitemShut {NoStop}%
\bibitem [{\citenamefont {Loredo}\ \emph {et~al.}(2016)\citenamefont {Loredo},
  \citenamefont {Zakaria}, \citenamefont {Somaschi}, \citenamefont {Anton},
  \citenamefont {de~Santis}, \citenamefont {Giesz}, \citenamefont {Grange},
  \citenamefont {Broome}, \citenamefont {Gazzano}, \citenamefont {Coppola},
  \citenamefont {Sagnes}, \citenamefont {Lemaitre}, \citenamefont {Auffeves},
  \citenamefont {Senellart}, \citenamefont {Almeida},\ and\ \citenamefont
  {White}}]{Loredo2016}%
  \BibitemOpen
  \bibfield  {author} {\bibinfo {author} {\bibfnamefont {J.~C.}\ \bibnamefont
  {Loredo}}, \bibinfo {author} {\bibfnamefont {N.~A.}\ \bibnamefont {Zakaria}},
  \bibinfo {author} {\bibfnamefont {N.}~\bibnamefont {Somaschi}}, \bibinfo
  {author} {\bibfnamefont {C.}~\bibnamefont {Anton}}, \bibinfo {author}
  {\bibfnamefont {L.}~\bibnamefont {de~Santis}}, \bibinfo {author}
  {\bibfnamefont {V.}~\bibnamefont {Giesz}}, \bibinfo {author} {\bibfnamefont
  {T.}~\bibnamefont {Grange}}, \bibinfo {author} {\bibfnamefont {M.~A.}\
  \bibnamefont {Broome}}, \bibinfo {author} {\bibfnamefont {O.}~\bibnamefont
  {Gazzano}}, \bibinfo {author} {\bibfnamefont {G.}~\bibnamefont {Coppola}},
  \bibinfo {author} {\bibfnamefont {I.}~\bibnamefont {Sagnes}}, \bibinfo
  {author} {\bibfnamefont {A.}~\bibnamefont {Lemaitre}}, \bibinfo {author}
  {\bibfnamefont {A.}~\bibnamefont {Auffeves}}, \bibinfo {author}
  {\bibfnamefont {P.}~\bibnamefont {Senellart}}, \bibinfo {author}
  {\bibfnamefont {M.~P.}\ \bibnamefont {Almeida}},\ and\ \bibinfo {author}
  {\bibfnamefont {A.~G.}\ \bibnamefont {White}},\ }\bibfield  {title} {\bibinfo
  {title} {{Scalable performance in solid-state single-photon sources}},\
  }\href {https://doi.org/10.1364/optica.3.000433} {\bibfield  {journal}
  {\bibinfo  {journal} {Optica}\ }\textbf {\bibinfo {volume} {3}},\ \bibinfo
  {pages} {433} (\bibinfo {year} {2016})}\BibitemShut {NoStop}%
\bibitem [{\citenamefont {Aharonovich}\ \emph {et~al.}(2016)\citenamefont
  {Aharonovich}, \citenamefont {Englund},\ and\ \citenamefont
  {Toth}}]{Aharonovich2016}%
  \BibitemOpen
  \bibfield  {author} {\bibinfo {author} {\bibfnamefont {I.}~\bibnamefont
  {Aharonovich}}, \bibinfo {author} {\bibfnamefont {D.}~\bibnamefont
  {Englund}},\ and\ \bibinfo {author} {\bibfnamefont {M.}~\bibnamefont
  {Toth}},\ }\bibfield  {title} {\bibinfo {title} {{Solid-state single-photon
  emitters}},\ }\href {https://doi.org/10.1038/nphoton.2016.186} {\bibfield
  {journal} {\bibinfo  {journal} {Nature Photonics}\ }\textbf {\bibinfo
  {volume} {10}},\ \bibinfo {pages} {631} (\bibinfo {year} {2016})}\BibitemShut
  {NoStop}%
\bibitem [{\citenamefont {Wolfowicz}\ \emph {et~al.}(2021)\citenamefont
  {Wolfowicz}, \citenamefont {Heremans}, \citenamefont {Anderson},
  \citenamefont {Kanai}, \citenamefont {Seo}, \citenamefont {Gali},
  \citenamefont {Galli},\ and\ \citenamefont {Awschalom}}]{Wolfowicz2021}%
  \BibitemOpen
  \bibfield  {author} {\bibinfo {author} {\bibfnamefont {G.}~\bibnamefont
  {Wolfowicz}}, \bibinfo {author} {\bibfnamefont {F.~J.}\ \bibnamefont
  {Heremans}}, \bibinfo {author} {\bibfnamefont {C.~P.}\ \bibnamefont
  {Anderson}}, \bibinfo {author} {\bibfnamefont {S.}~\bibnamefont {Kanai}},
  \bibinfo {author} {\bibfnamefont {H.}~\bibnamefont {Seo}}, \bibinfo {author}
  {\bibfnamefont {A.}~\bibnamefont {Gali}}, \bibinfo {author} {\bibfnamefont
  {G.}~\bibnamefont {Galli}},\ and\ \bibinfo {author} {\bibfnamefont {D.~D.}\
  \bibnamefont {Awschalom}},\ }\bibfield  {title} {\bibinfo {title} {{Quantum
  guidelines for solid-state spin defects}},\ }\href
  {https://doi.org/10.1038/s41578-021-00306-y} {\bibfield  {journal} {\bibinfo
  {journal} {Nature Reviews Materials}\ }\textbf {\bibinfo {volume} {10}},\
  \bibinfo {pages} {906} (\bibinfo {year} {2021})}\BibitemShut {NoStop}%
\bibitem [{\citenamefont {Tran}\ \emph {et~al.}(2016)\citenamefont {Tran},
  \citenamefont {Bray}, \citenamefont {Ford}, \citenamefont {Toth},\ and\
  \citenamefont {Aharonovich}}]{Tran2016}%
  \BibitemOpen
  \bibfield  {author} {\bibinfo {author} {\bibfnamefont {T.~T.}\ \bibnamefont
  {Tran}}, \bibinfo {author} {\bibfnamefont {K.}~\bibnamefont {Bray}}, \bibinfo
  {author} {\bibfnamefont {M.~J.}\ \bibnamefont {Ford}}, \bibinfo {author}
  {\bibfnamefont {M.}~\bibnamefont {Toth}},\ and\ \bibinfo {author}
  {\bibfnamefont {I.}~\bibnamefont {Aharonovich}},\ }\bibfield  {title}
  {\bibinfo {title} {{Quantum emission from hexagonal boron nitride
  monolayers}},\ }\href {https://doi.org/10.1038/nnano.2015.242} {\bibfield
  {journal} {\bibinfo  {journal} {Nature Nanotechnology}\ }\textbf {\bibinfo
  {volume} {11}},\ \bibinfo {pages} {37} (\bibinfo {year} {2016})}\BibitemShut
  {NoStop}%
\bibitem [{\citenamefont {Koperski}\ \emph {et~al.}(2018)\citenamefont
  {Koperski}, \citenamefont {Nogajewski},\ and\ \citenamefont
  {Potemski}}]{Koperski2018}%
  \BibitemOpen
  \bibfield  {author} {\bibinfo {author} {\bibfnamefont {M.}~\bibnamefont
  {Koperski}}, \bibinfo {author} {\bibfnamefont {K.}~\bibnamefont
  {Nogajewski}},\ and\ \bibinfo {author} {\bibfnamefont {M.}~\bibnamefont
  {Potemski}},\ }\bibfield  {title} {\bibinfo {title} {{Single photon emitters
  in boron nitride: More than a supplementary material}},\ }\href
  {https://doi.org/10.1016/j.optcom.2017.10.083} {\bibfield  {journal}
  {\bibinfo  {journal} {Optics Communications}\ }\textbf {\bibinfo {volume}
  {411}},\ \bibinfo {pages} {158} (\bibinfo {year} {2018})}\BibitemShut
  {NoStop}%
\bibitem [{\citenamefont {Exarhos}\ \emph {et~al.}(2019)\citenamefont
  {Exarhos}, \citenamefont {Hopper}, \citenamefont {Patel}, \citenamefont
  {Doherty},\ and\ \citenamefont {Bassett}}]{Exarhos2019}%
  \BibitemOpen
  \bibfield  {author} {\bibinfo {author} {\bibfnamefont {A.~L.}\ \bibnamefont
  {Exarhos}}, \bibinfo {author} {\bibfnamefont {D.~A.}\ \bibnamefont {Hopper}},
  \bibinfo {author} {\bibfnamefont {R.~N.}\ \bibnamefont {Patel}}, \bibinfo
  {author} {\bibfnamefont {M.~W.}\ \bibnamefont {Doherty}},\ and\ \bibinfo
  {author} {\bibfnamefont {L.~C.}\ \bibnamefont {Bassett}},\ }\bibfield
  {title} {\bibinfo {title} {{Magnetic-field-dependent quantum emission in
  hexagonal boron nitride at room temperature}},\ }\href
  {https://doi.org/10.1038/s41467-018-08185-8} {\bibfield  {journal} {\bibinfo
  {journal} {Nature Communications}\ }\textbf {\bibinfo {volume} {10}},\
  \bibinfo {pages} {1} (\bibinfo {year} {2019})}\BibitemShut {NoStop}%
\bibitem [{\citenamefont {Stern}\ \emph {et~al.}(2019)\citenamefont {Stern},
  \citenamefont {Wang}, \citenamefont {Fan}, \citenamefont {Mizuta},
  \citenamefont {Stewart}, \citenamefont {Needham}, \citenamefont {Roberts},
  \citenamefont {Wai}, \citenamefont {Ginsberg}, \citenamefont {Klenerman},
  \citenamefont {Hofmann},\ and\ \citenamefont {Lee}}]{Stern2019}%
  \BibitemOpen
  \bibfield  {author} {\bibinfo {author} {\bibfnamefont {H.~L.}\ \bibnamefont
  {Stern}}, \bibinfo {author} {\bibfnamefont {R.}~\bibnamefont {Wang}},
  \bibinfo {author} {\bibfnamefont {Y.}~\bibnamefont {Fan}}, \bibinfo {author}
  {\bibfnamefont {R.}~\bibnamefont {Mizuta}}, \bibinfo {author} {\bibfnamefont
  {J.~C.}\ \bibnamefont {Stewart}}, \bibinfo {author} {\bibfnamefont {L.~M.}\
  \bibnamefont {Needham}}, \bibinfo {author} {\bibfnamefont {T.~D.}\
  \bibnamefont {Roberts}}, \bibinfo {author} {\bibfnamefont {R.}~\bibnamefont
  {Wai}}, \bibinfo {author} {\bibfnamefont {N.~S.}\ \bibnamefont {Ginsberg}},
  \bibinfo {author} {\bibfnamefont {D.}~\bibnamefont {Klenerman}}, \bibinfo
  {author} {\bibfnamefont {S.}~\bibnamefont {Hofmann}},\ and\ \bibinfo {author}
  {\bibfnamefont {S.~F.}\ \bibnamefont {Lee}},\ }\bibfield  {title} {\bibinfo
  {title} {{Spectrally Resolved Photodynamics of Individual Emitters in
  Large-Area Monolayers of Hexagonal Boron Nitride}},\ }\href
  {https://doi.org/10.1021/acsnano.9b00274} {\bibfield  {journal} {\bibinfo
  {journal} {ACS Nano}\ }\textbf {\bibinfo {volume} {13}},\ \bibinfo {pages}
  {4538} (\bibinfo {year} {2019})}\BibitemShut {NoStop}%
\bibitem [{\citenamefont {Reserbat-Plantey}\ \emph {et~al.}(2021)\citenamefont
  {Reserbat-Plantey}, \citenamefont {Epstein}, \citenamefont {Torre},
  \citenamefont {Costa}, \citenamefont {Gon{\c{c}}alves}, \citenamefont
  {Mortensen}, \citenamefont {Polini}, \citenamefont {Song}, \citenamefont
  {Peres},\ and\ \citenamefont {Koppens}}]{Reserbat-Plantey2021}%
  \BibitemOpen
  \bibfield  {author} {\bibinfo {author} {\bibfnamefont {A.}~\bibnamefont
  {Reserbat-Plantey}}, \bibinfo {author} {\bibfnamefont {I.}~\bibnamefont
  {Epstein}}, \bibinfo {author} {\bibfnamefont {I.}~\bibnamefont {Torre}},
  \bibinfo {author} {\bibfnamefont {A.~T.}\ \bibnamefont {Costa}}, \bibinfo
  {author} {\bibfnamefont {P.~A.}\ \bibnamefont {Gon{\c{c}}alves}}, \bibinfo
  {author} {\bibfnamefont {N.~A.}\ \bibnamefont {Mortensen}}, \bibinfo {author}
  {\bibfnamefont {M.}~\bibnamefont {Polini}}, \bibinfo {author} {\bibfnamefont
  {J.~C.}\ \bibnamefont {Song}}, \bibinfo {author} {\bibfnamefont {N.~M.}\
  \bibnamefont {Peres}},\ and\ \bibinfo {author} {\bibfnamefont {F.~H.}\
  \bibnamefont {Koppens}},\ }\bibfield  {title} {\bibinfo {title} {{Quantum
  Nanophotonics in Two-Dimensional Materials}},\ }\href
  {https://doi.org/10.1021/acsphotonics.0c01224} {\bibfield  {journal}
  {\bibinfo  {journal} {ACS Photonics}\ }\textbf {\bibinfo {volume} {8}},\
  \bibinfo {pages} {85} (\bibinfo {year} {2021})}\BibitemShut {NoStop}%
\bibitem [{\citenamefont {Azzam}\ \emph {et~al.}(2021)\citenamefont {Azzam},
  \citenamefont {Parto},\ and\ \citenamefont {Moody}}]{Azzam2021}%
  \BibitemOpen
  \bibfield  {author} {\bibinfo {author} {\bibfnamefont {S.~I.}\ \bibnamefont
  {Azzam}}, \bibinfo {author} {\bibfnamefont {K.}~\bibnamefont {Parto}},\ and\
  \bibinfo {author} {\bibfnamefont {G.}~\bibnamefont {Moody}},\ }\bibfield
  {title} {\bibinfo {title} {{Prospects and challenges of quantum emitters in
  2D materials}},\ }\href {https://doi.org/10.1063/5.0054116} {\bibfield
  {journal} {\bibinfo  {journal} {Applied Physics Letters}\ }\textbf {\bibinfo
  {volume} {118}},\ \bibinfo {pages} {240502} (\bibinfo {year}
  {2021})}\BibitemShut {NoStop}%
\bibitem [{\citenamefont {Palacios-Berraquero}\ \emph
  {et~al.}(2017)\citenamefont {Palacios-Berraquero}, \citenamefont {Kara},
  \citenamefont {Montblanch}, \citenamefont {Barbone}, \citenamefont
  {Latawiec}, \citenamefont {Yoon}, \citenamefont {Ott}, \citenamefont
  {Loncar}, \citenamefont {Ferrari},\ and\ \citenamefont
  {Atat{\"{u}}re}}]{Palacios-Berraquero2017}%
  \BibitemOpen
  \bibfield  {author} {\bibinfo {author} {\bibfnamefont {C.}~\bibnamefont
  {Palacios-Berraquero}}, \bibinfo {author} {\bibfnamefont {D.~M.}\
  \bibnamefont {Kara}}, \bibinfo {author} {\bibfnamefont {A.~R.}\ \bibnamefont
  {Montblanch}}, \bibinfo {author} {\bibfnamefont {M.}~\bibnamefont {Barbone}},
  \bibinfo {author} {\bibfnamefont {P.}~\bibnamefont {Latawiec}}, \bibinfo
  {author} {\bibfnamefont {D.}~\bibnamefont {Yoon}}, \bibinfo {author}
  {\bibfnamefont {A.~K.}\ \bibnamefont {Ott}}, \bibinfo {author} {\bibfnamefont
  {M.}~\bibnamefont {Loncar}}, \bibinfo {author} {\bibfnamefont {A.~C.}\
  \bibnamefont {Ferrari}},\ and\ \bibinfo {author} {\bibfnamefont
  {M.}~\bibnamefont {Atat{\"{u}}re}},\ }\bibfield  {title} {\bibinfo {title}
  {{Large-scale quantum-emitter arrays in atomically thin semiconductors}},\
  }\href {https://doi.org/10.1038/ncomms15093} {\bibfield  {journal} {\bibinfo
  {journal} {Nature Communications}\ }\textbf {\bibinfo {volume} {8}},\
  \bibinfo {pages} {1} (\bibinfo {year} {2017})}\BibitemShut {NoStop}%
\bibitem [{\citenamefont {Chakraborty}\ \emph {et~al.}(2019)\citenamefont
  {Chakraborty}, \citenamefont {Lehmann}, \citenamefont {Zhang}, \citenamefont
  {Borgsdorf}, \citenamefont {W{\"{o}}hrl}, \citenamefont {Remfort},
  \citenamefont {Buck}, \citenamefont {K{\"{o}}hler},\ and\ \citenamefont
  {Suter}}]{Chakraborty2019}%
  \BibitemOpen
  \bibfield  {author} {\bibinfo {author} {\bibfnamefont {T.}~\bibnamefont
  {Chakraborty}}, \bibinfo {author} {\bibfnamefont {F.}~\bibnamefont
  {Lehmann}}, \bibinfo {author} {\bibfnamefont {J.}~\bibnamefont {Zhang}},
  \bibinfo {author} {\bibfnamefont {S.}~\bibnamefont {Borgsdorf}}, \bibinfo
  {author} {\bibfnamefont {N.}~\bibnamefont {W{\"{o}}hrl}}, \bibinfo {author}
  {\bibfnamefont {R.}~\bibnamefont {Remfort}}, \bibinfo {author} {\bibfnamefont
  {V.}~\bibnamefont {Buck}}, \bibinfo {author} {\bibfnamefont {U.}~\bibnamefont
  {K{\"{o}}hler}},\ and\ \bibinfo {author} {\bibfnamefont {D.}~\bibnamefont
  {Suter}},\ }\bibfield  {title} {\bibinfo {title} {{CVD growth of ultrapure
  diamond, generation of NV centers by ion implantation, and their
  spectroscopic characterization for quantum technological applications}},\
  }\href {https://doi.org/10.1103/PhysRevMaterials.3.065205} {\bibfield
  {journal} {\bibinfo  {journal} {Physical Review Materials}\ }\textbf
  {\bibinfo {volume} {3}},\ \bibinfo {pages} {065205} (\bibinfo {year}
  {2019})}\BibitemShut {NoStop}%
\bibitem [{\citenamefont {Dang}\ \emph {et~al.}(2020)\citenamefont {Dang},
  \citenamefont {Sun}, \citenamefont {Xie}, \citenamefont {Yu}, \citenamefont
  {Peng}, \citenamefont {Qian}, \citenamefont {Wu}, \citenamefont {Song},
  \citenamefont {Yang}, \citenamefont {Xiao}, \citenamefont {Yang},
  \citenamefont {Wang}, \citenamefont {Rafiq}, \citenamefont {Wang},\ and\
  \citenamefont {Xu}}]{Dang2020}%
  \BibitemOpen
  \bibfield  {author} {\bibinfo {author} {\bibfnamefont {J.}~\bibnamefont
  {Dang}}, \bibinfo {author} {\bibfnamefont {S.}~\bibnamefont {Sun}}, \bibinfo
  {author} {\bibfnamefont {X.}~\bibnamefont {Xie}}, \bibinfo {author}
  {\bibfnamefont {Y.}~\bibnamefont {Yu}}, \bibinfo {author} {\bibfnamefont
  {K.}~\bibnamefont {Peng}}, \bibinfo {author} {\bibfnamefont {C.}~\bibnamefont
  {Qian}}, \bibinfo {author} {\bibfnamefont {S.}~\bibnamefont {Wu}}, \bibinfo
  {author} {\bibfnamefont {F.}~\bibnamefont {Song}}, \bibinfo {author}
  {\bibfnamefont {J.}~\bibnamefont {Yang}}, \bibinfo {author} {\bibfnamefont
  {S.}~\bibnamefont {Xiao}}, \bibinfo {author} {\bibfnamefont {L.}~\bibnamefont
  {Yang}}, \bibinfo {author} {\bibfnamefont {Y.}~\bibnamefont {Wang}}, \bibinfo
  {author} {\bibfnamefont {M.~A.}\ \bibnamefont {Rafiq}}, \bibinfo {author}
  {\bibfnamefont {C.}~\bibnamefont {Wang}},\ and\ \bibinfo {author}
  {\bibfnamefont {X.}~\bibnamefont {Xu}},\ }\bibfield  {title} {\bibinfo
  {title} {{Identifying defect-related quantum emitters in monolayer WSe2}},\
  }\href {https://doi.org/10.1038/s41699-020-0136-0} {\bibfield  {journal}
  {\bibinfo  {journal} {npj 2D Materials and Applications}\ }\textbf {\bibinfo
  {volume} {4}},\ \bibinfo {pages} {1} (\bibinfo {year} {2020})}\BibitemShut
  {NoStop}%
\bibitem [{\citenamefont {Bassett}\ \emph {et~al.}(2019)\citenamefont
  {Bassett}, \citenamefont {Alkauskas}, \citenamefont {Exarhos},\ and\
  \citenamefont {Fu}}]{Bassett2019}%
  \BibitemOpen
  \bibfield  {author} {\bibinfo {author} {\bibfnamefont {L.~C.}\ \bibnamefont
  {Bassett}}, \bibinfo {author} {\bibfnamefont {A.}~\bibnamefont {Alkauskas}},
  \bibinfo {author} {\bibfnamefont {A.~L.}\ \bibnamefont {Exarhos}},\ and\
  \bibinfo {author} {\bibfnamefont {K.~M.~C.}\ \bibnamefont {Fu}},\ }\bibfield
  {title} {\bibinfo {title} {{Quantum defects by design}},\ }\href
  {https://doi.org/10.1515/nanoph-2019-0211} {\bibfield  {journal} {\bibinfo
  {journal} {Nanophotonics}\ }\textbf {\bibinfo {volume} {8}},\ \bibinfo
  {pages} {1867} (\bibinfo {year} {2019})}\BibitemShut {NoStop}%
\bibitem [{\citenamefont {Shenoy}\ \emph {et~al.}(2020)\citenamefont {Shenoy},
  \citenamefont {Frey}, \citenamefont {Akinwande},\ and\ \citenamefont
  {Jariwala}}]{Shenoy2020}%
  \BibitemOpen
  \bibfield  {author} {\bibinfo {author} {\bibfnamefont {V.~B.}\ \bibnamefont
  {Shenoy}}, \bibinfo {author} {\bibfnamefont {N.~C.}\ \bibnamefont {Frey}},
  \bibinfo {author} {\bibfnamefont {D.}~\bibnamefont {Akinwande}},\ and\
  \bibinfo {author} {\bibfnamefont {D.}~\bibnamefont {Jariwala}},\ }\bibfield
  {title} {\bibinfo {title} {{Machine learning-enabled design of point defects
  in 2d materials for quantum and neuromorphic information processing}},\
  }\href {https://doi.org/10.1021/acsnano.0c05267} {\bibfield  {journal}
  {\bibinfo  {journal} {ACS Nano}\ }\textbf {\bibinfo {volume} {14}},\ \bibinfo
  {pages} {13406} (\bibinfo {year} {2020})}\BibitemShut {NoStop}%
\bibitem [{\citenamefont {Ferrenti}\ \emph {et~al.}(2020)\citenamefont
  {Ferrenti}, \citenamefont {de~Leon}, \citenamefont {Thompson},\ and\
  \citenamefont {Cava}}]{Ferrenti2020}%
  \BibitemOpen
  \bibfield  {author} {\bibinfo {author} {\bibfnamefont {A.~M.}\ \bibnamefont
  {Ferrenti}}, \bibinfo {author} {\bibfnamefont {N.~P.}\ \bibnamefont
  {de~Leon}}, \bibinfo {author} {\bibfnamefont {J.~D.}\ \bibnamefont
  {Thompson}},\ and\ \bibinfo {author} {\bibfnamefont {R.~J.}\ \bibnamefont
  {Cava}},\ }\bibfield  {title} {\bibinfo {title} {{Identifying candidate hosts
  for quantum defects via data mining}},\ }\href
  {https://doi.org/10.1038/s41524-020-00391-7} {\bibfield  {journal} {\bibinfo
  {journal} {npj Computational Materials}\ }\textbf {\bibinfo {volume} {6}},\
  \bibinfo {pages} {1} (\bibinfo {year} {2020})}\BibitemShut {NoStop}%
\bibitem [{\citenamefont {Stewart}\ \emph {et~al.}(2019)\citenamefont
  {Stewart}, \citenamefont {Kianinia}, \citenamefont {Previdi}, \citenamefont
  {Tran}, \citenamefont {Aharonovich},\ and\ \citenamefont
  {Bradac}}]{Stewart2019}%
  \BibitemOpen
  \bibfield  {author} {\bibinfo {author} {\bibfnamefont {C.}~\bibnamefont
  {Stewart}}, \bibinfo {author} {\bibfnamefont {M.}~\bibnamefont {Kianinia}},
  \bibinfo {author} {\bibfnamefont {R.}~\bibnamefont {Previdi}}, \bibinfo
  {author} {\bibfnamefont {T.~T.}\ \bibnamefont {Tran}}, \bibinfo {author}
  {\bibfnamefont {I.}~\bibnamefont {Aharonovich}},\ and\ \bibinfo {author}
  {\bibfnamefont {C.}~\bibnamefont {Bradac}},\ }\bibfield  {title} {\bibinfo
  {title} {{Quantum emission from localized defects in zinc sulfide}},\ }\href
  {https://doi.org/10.1364/ol.44.004873} {\bibfield  {journal} {\bibinfo
  {journal} {Optics Letters}\ }\textbf {\bibinfo {volume} {44}},\ \bibinfo
  {pages} {4873} (\bibinfo {year} {2019})}\BibitemShut {NoStop}%
\bibitem [{\citenamefont {Linpeng}\ \emph {et~al.}(2018)\citenamefont
  {Linpeng}, \citenamefont {Viitaniemi}, \citenamefont {Vishnuradhan},
  \citenamefont {Kozuka}, \citenamefont {Johnson}, \citenamefont {Kawasaki},\
  and\ \citenamefont {Fu}}]{Linpeng2018}%
  \BibitemOpen
  \bibfield  {author} {\bibinfo {author} {\bibfnamefont {X.}~\bibnamefont
  {Linpeng}}, \bibinfo {author} {\bibfnamefont {M.~L.}\ \bibnamefont
  {Viitaniemi}}, \bibinfo {author} {\bibfnamefont {A.}~\bibnamefont
  {Vishnuradhan}}, \bibinfo {author} {\bibfnamefont {Y.}~\bibnamefont
  {Kozuka}}, \bibinfo {author} {\bibfnamefont {C.}~\bibnamefont {Johnson}},
  \bibinfo {author} {\bibfnamefont {M.}~\bibnamefont {Kawasaki}},\ and\
  \bibinfo {author} {\bibfnamefont {K.~M.~C.}\ \bibnamefont {Fu}},\ }\bibfield
  {title} {\bibinfo {title} {{Coherence Properties of Shallow Donor Qubits in
  Zn O}},\ }\href {https://doi.org/10.1103/PhysRevApplied.10.064061} {\bibfield
   {journal} {\bibinfo  {journal} {Physical Review Applied}\ }\textbf {\bibinfo
  {volume} {10}},\ \bibinfo {pages} {064061} (\bibinfo {year}
  {2018})}\BibitemShut {NoStop}%
\bibitem [{\citenamefont {Chung}\ \emph {et~al.}(2018)\citenamefont {Chung},
  \citenamefont {Leung}, \citenamefont {To}, \citenamefont
  {Djuri{\v{s}}i{\'{c}}},\ and\ \citenamefont {Tomljenovic-Hanic}}]{Chung2018}%
  \BibitemOpen
  \bibfield  {author} {\bibinfo {author} {\bibfnamefont {K.}~\bibnamefont
  {Chung}}, \bibinfo {author} {\bibfnamefont {Y.~H.}\ \bibnamefont {Leung}},
  \bibinfo {author} {\bibfnamefont {C.~H.}\ \bibnamefont {To}}, \bibinfo
  {author} {\bibfnamefont {A.~B.}\ \bibnamefont {Djuri{\v{s}}i{\'{c}}}},\ and\
  \bibinfo {author} {\bibfnamefont {S.}~\bibnamefont {Tomljenovic-Hanic}},\
  }\bibfield  {title} {\bibinfo {title} {{Room-temperature single-photon
  emitters in titanium dioxide optical defects}},\ }\href
  {https://doi.org/10.3762/bjnano.9.100} {\bibfield  {journal} {\bibinfo
  {journal} {Beilstein Journal of Nanotechnology}\ }\textbf {\bibinfo {volume}
  {9}},\ \bibinfo {pages} {1085} (\bibinfo {year} {2018})}\BibitemShut
  {NoStop}%
\bibitem [{\citenamefont {Berhane}\ \emph {et~al.}(2017)\citenamefont
  {Berhane}, \citenamefont {Jeong}, \citenamefont {Bodrog}, \citenamefont
  {Fiedler}, \citenamefont {Schr{\"{o}}der}, \citenamefont {Trivi{\~{n}}o},
  \citenamefont {Palacios}, \citenamefont {Gali}, \citenamefont {Toth},
  \citenamefont {Englund},\ and\ \citenamefont {Aharonovich}}]{Berhane2017}%
  \BibitemOpen
  \bibfield  {author} {\bibinfo {author} {\bibfnamefont {A.~M.}\ \bibnamefont
  {Berhane}}, \bibinfo {author} {\bibfnamefont {K.~Y.}\ \bibnamefont {Jeong}},
  \bibinfo {author} {\bibfnamefont {Z.}~\bibnamefont {Bodrog}}, \bibinfo
  {author} {\bibfnamefont {S.}~\bibnamefont {Fiedler}}, \bibinfo {author}
  {\bibfnamefont {T.}~\bibnamefont {Schr{\"{o}}der}}, \bibinfo {author}
  {\bibfnamefont {N.~V.}\ \bibnamefont {Trivi{\~{n}}o}}, \bibinfo {author}
  {\bibfnamefont {T.}~\bibnamefont {Palacios}}, \bibinfo {author}
  {\bibfnamefont {A.}~\bibnamefont {Gali}}, \bibinfo {author} {\bibfnamefont
  {M.}~\bibnamefont {Toth}}, \bibinfo {author} {\bibfnamefont {D.}~\bibnamefont
  {Englund}},\ and\ \bibinfo {author} {\bibfnamefont {I.}~\bibnamefont
  {Aharonovich}},\ }\bibfield  {title} {\bibinfo {title} {{Bright
  Room-Temperature Single-Photon Emission from Defects in Gallium Nitride}},\
  }\href {https://doi.org/10.1002/adma.201605092} {\bibfield  {journal}
  {\bibinfo  {journal} {Advanced Materials}\ }\textbf {\bibinfo {volume}
  {29}},\ \bibinfo {pages} {1605092} (\bibinfo {year} {2017})}\BibitemShut
  {NoStop}%
\bibitem [{\citenamefont {Kagan}\ \emph {et~al.}(2021)\citenamefont {Kagan},
  \citenamefont {Bassett}, \citenamefont {Murray},\ and\ \citenamefont
  {Thompson}}]{Kagan2021}%
  \BibitemOpen
  \bibfield  {author} {\bibinfo {author} {\bibfnamefont {C.~R.}\ \bibnamefont
  {Kagan}}, \bibinfo {author} {\bibfnamefont {L.~C.}\ \bibnamefont {Bassett}},
  \bibinfo {author} {\bibfnamefont {C.~B.}\ \bibnamefont {Murray}},\ and\
  \bibinfo {author} {\bibfnamefont {S.~M.}\ \bibnamefont {Thompson}},\
  }\bibfield  {title} {\bibinfo {title} {{Colloidal Quantum Dots as Platforms
  for Quantum Information Science}},\ }\href
  {https://doi.org/10.1021/acs.chemrev.0c00831} {\bibfield  {journal} {\bibinfo
   {journal} {Chemical Reviews}\ }\textbf {\bibinfo {volume} {121}},\ \bibinfo
  {pages} {3186} (\bibinfo {year} {2021})}\BibitemShut {NoStop}%
\bibitem [{\citenamefont {Wang}\ \emph {et~al.}(2006)\citenamefont {Wang},
  \citenamefont {Kurtsiefer}, \citenamefont {Weinfurter},\ and\ \citenamefont
  {Burchard}}]{Wang2006}%
  \BibitemOpen
  \bibfield  {author} {\bibinfo {author} {\bibfnamefont {C.}~\bibnamefont
  {Wang}}, \bibinfo {author} {\bibfnamefont {C.}~\bibnamefont {Kurtsiefer}},
  \bibinfo {author} {\bibfnamefont {H.}~\bibnamefont {Weinfurter}},\ and\
  \bibinfo {author} {\bibfnamefont {B.}~\bibnamefont {Burchard}},\ }\bibfield
  {title} {\bibinfo {title} {{Single photon emission from SiV centres in
  diamond produced by ion implantation}},\ }\href
  {https://doi.org/10.1088/0953-4075/39/1/005} {\bibfield  {journal} {\bibinfo
  {journal} {J. Phys. B -- At. Mol. Opt.}\ }\textbf {\bibinfo {volume} {39}},\
  \bibinfo {pages} {37} (\bibinfo {year} {2006})}\BibitemShut {NoStop}%
\bibitem [{\citenamefont {Toyli}\ \emph {et~al.}(2010)\citenamefont {Toyli},
  \citenamefont {Weis}, \citenamefont {Fuchs}, \citenamefont {Schenkel},\ and\
  \citenamefont {Awschalom}}]{Toyli2010}%
  \BibitemOpen
  \bibfield  {author} {\bibinfo {author} {\bibfnamefont {D.~M.}\ \bibnamefont
  {Toyli}}, \bibinfo {author} {\bibfnamefont {C.~D.}\ \bibnamefont {Weis}},
  \bibinfo {author} {\bibfnamefont {G.~D.}\ \bibnamefont {Fuchs}}, \bibinfo
  {author} {\bibfnamefont {T.}~\bibnamefont {Schenkel}},\ and\ \bibinfo
  {author} {\bibfnamefont {D.~D.}\ \bibnamefont {Awschalom}},\ }\bibfield
  {title} {\bibinfo {title} {{Chip-scale nanofabrication of single spins and
  spin arrays in diamond}},\ }\href {https://doi.org/10.1021/nl102066q}
  {\bibfield  {journal} {\bibinfo  {journal} {Nano Letters}\ }\textbf {\bibinfo
  {volume} {10}},\ \bibinfo {pages} {3168} (\bibinfo {year}
  {2010})}\BibitemShut {NoStop}%
\bibitem [{\citenamefont {Hollenbach}\ \emph {et~al.}(2020)\citenamefont
  {Hollenbach}, \citenamefont {Berenc{\'{e}}n}, \citenamefont {Kentsch},
  \citenamefont {Helm},\ and\ \citenamefont {Astakhov}}]{Hollenbach2020}%
  \BibitemOpen
  \bibfield  {author} {\bibinfo {author} {\bibfnamefont {M.}~\bibnamefont
  {Hollenbach}}, \bibinfo {author} {\bibfnamefont {Y.}~\bibnamefont
  {Berenc{\'{e}}n}}, \bibinfo {author} {\bibfnamefont {U.}~\bibnamefont
  {Kentsch}}, \bibinfo {author} {\bibfnamefont {M.}~\bibnamefont {Helm}},\ and\
  \bibinfo {author} {\bibfnamefont {G.~V.}\ \bibnamefont {Astakhov}},\
  }\bibfield  {title} {\bibinfo {title} {{Engineering telecom single-photon
  emitters in silicon for scalable quantum photonics}},\ }\href
  {https://doi.org/10.1364/oe.397377} {\bibfield  {journal} {\bibinfo
  {journal} {Optics Express}\ }\textbf {\bibinfo {volume} {28}},\ \bibinfo
  {pages} {26111} (\bibinfo {year} {2020})}\BibitemShut {NoStop}%
\bibitem [{\citenamefont {Rodt}\ \emph {et~al.}(2020)\citenamefont {Rodt},
  \citenamefont {Reitzenstein},\ and\ \citenamefont {Heindel}}]{Rodt2020}%
  \BibitemOpen
  \bibfield  {author} {\bibinfo {author} {\bibfnamefont {S.}~\bibnamefont
  {Rodt}}, \bibinfo {author} {\bibfnamefont {S.}~\bibnamefont {Reitzenstein}},\
  and\ \bibinfo {author} {\bibfnamefont {T.}~\bibnamefont {Heindel}},\
  }\bibfield  {title} {\bibinfo {title} {{Deterministically fabricated
  solid-state quantum-light sources}},\ }\href
  {https://doi.org/10.1088/1361-648X/ab5e15} {\bibfield  {journal} {\bibinfo
  {journal} {Journal of Physics Condensed Matter}\ }\textbf {\bibinfo {volume}
  {32}},\ \bibinfo {pages} {153003} (\bibinfo {year} {2020})}\BibitemShut
  {NoStop}%
\bibitem [{\citenamefont {Klein}\ \emph {et~al.}(2020)\citenamefont {Klein},
  \citenamefont {Sigl}, \citenamefont {H{\"{o}}tger}, \citenamefont {Gyger},
  \citenamefont {Barthelmi}, \citenamefont {Florian}, \citenamefont {Kerelsky},
  \citenamefont {Mitterreiter}, \citenamefont {Kastl}, \citenamefont {Rey},
  \citenamefont {Taniguchi}, \citenamefont {Watanabe}, \citenamefont {Jahnke},
  \citenamefont {Zwiller}, \citenamefont {J{\"{o}}ns}, \citenamefont
  {Pasupathy}, \citenamefont {Ross}, \citenamefont {M{\"{u}}ller},
  \citenamefont {Wurstbauer}, \citenamefont {Finley},\ and\ \citenamefont
  {Holleitner}}]{Klein2020}%
  \BibitemOpen
  \bibfield  {author} {\bibinfo {author} {\bibfnamefont {J.}~\bibnamefont
  {Klein}}, \bibinfo {author} {\bibfnamefont {L.}~\bibnamefont {Sigl}},
  \bibinfo {author} {\bibfnamefont {A.}~\bibnamefont {H{\"{o}}tger}}, \bibinfo
  {author} {\bibfnamefont {S.}~\bibnamefont {Gyger}}, \bibinfo {author}
  {\bibfnamefont {K.}~\bibnamefont {Barthelmi}}, \bibinfo {author}
  {\bibfnamefont {M.}~\bibnamefont {Florian}}, \bibinfo {author} {\bibfnamefont
  {A.}~\bibnamefont {Kerelsky}}, \bibinfo {author} {\bibfnamefont
  {E.}~\bibnamefont {Mitterreiter}}, \bibinfo {author} {\bibfnamefont
  {C.}~\bibnamefont {Kastl}}, \bibinfo {author} {\bibfnamefont
  {S.}~\bibnamefont {Rey}}, \bibinfo {author} {\bibfnamefont {T.}~\bibnamefont
  {Taniguchi}}, \bibinfo {author} {\bibfnamefont {K.}~\bibnamefont {Watanabe}},
  \bibinfo {author} {\bibfnamefont {F.}~\bibnamefont {Jahnke}}, \bibinfo
  {author} {\bibfnamefont {V.}~\bibnamefont {Zwiller}}, \bibinfo {author}
  {\bibfnamefont {K.~D.}\ \bibnamefont {J{\"{o}}ns}}, \bibinfo {author}
  {\bibfnamefont {A.}~\bibnamefont {Pasupathy}}, \bibinfo {author}
  {\bibfnamefont {F.~M.}\ \bibnamefont {Ross}}, \bibinfo {author}
  {\bibfnamefont {K.}~\bibnamefont {M{\"{u}}ller}}, \bibinfo {author}
  {\bibfnamefont {U.}~\bibnamefont {Wurstbauer}}, \bibinfo {author}
  {\bibfnamefont {J.~J.}\ \bibnamefont {Finley}},\ and\ \bibinfo {author}
  {\bibfnamefont {A.~W.}\ \bibnamefont {Holleitner}},\ }\bibfield  {title}
  {\bibinfo {title} {{Scalable single-photon sources in atomically thin
  {MoS}$_2$}},\ }\href@noop {} {\bibfield  {journal} {\bibinfo  {journal}
  {arXiv preprint arXiv:2002.08819}\ } (\bibinfo {year} {2020})},\ \Eprint
  {https://arxiv.org/abs/https://arxiv.org/abs/2002.08819}
  {https://arxiv.org/abs/2002.08819} \BibitemShut {NoStop}%
\bibitem [{\citenamefont {Lagomarsino}\ \emph {et~al.}(2021)\citenamefont
  {Lagomarsino}, \citenamefont {Flatae}, \citenamefont {Kambalathmana},
  \citenamefont {Sledz}, \citenamefont {Hunold}, \citenamefont {Soltani},
  \citenamefont {Reuschel}, \citenamefont {Sciortino}, \citenamefont {Gelli},
  \citenamefont {Massi}, \citenamefont {Czelusniak}, \citenamefont {Giuntini},\
  and\ \citenamefont {Agio}}]{Lagomarsino2021}%
  \BibitemOpen
  \bibfield  {author} {\bibinfo {author} {\bibfnamefont {S.}~\bibnamefont
  {Lagomarsino}}, \bibinfo {author} {\bibfnamefont {A.~M.}\ \bibnamefont
  {Flatae}}, \bibinfo {author} {\bibfnamefont {H.}~\bibnamefont
  {Kambalathmana}}, \bibinfo {author} {\bibfnamefont {F.}~\bibnamefont
  {Sledz}}, \bibinfo {author} {\bibfnamefont {L.}~\bibnamefont {Hunold}},
  \bibinfo {author} {\bibfnamefont {N.}~\bibnamefont {Soltani}}, \bibinfo
  {author} {\bibfnamefont {P.}~\bibnamefont {Reuschel}}, \bibinfo {author}
  {\bibfnamefont {S.}~\bibnamefont {Sciortino}}, \bibinfo {author}
  {\bibfnamefont {N.}~\bibnamefont {Gelli}}, \bibinfo {author} {\bibfnamefont
  {M.}~\bibnamefont {Massi}}, \bibinfo {author} {\bibfnamefont
  {C.}~\bibnamefont {Czelusniak}}, \bibinfo {author} {\bibfnamefont
  {L.}~\bibnamefont {Giuntini}},\ and\ \bibinfo {author} {\bibfnamefont
  {M.}~\bibnamefont {Agio}},\ }\bibfield  {title} {\bibinfo {title} {{Creation
  of Silicon-Vacancy Color Centers in Diamond by Ion Implantation}},\ }\href
  {https://doi.org/10.3389/fphy.2020.601362} {\bibfield  {journal} {\bibinfo
  {journal} {Frontiers in Physics}\ }\textbf {\bibinfo {volume} {8}},\ \bibinfo
  {pages} {626} (\bibinfo {year} {2021})}\BibitemShut {NoStop}%
\bibitem [{\citenamefont {Fuchs}\ \emph {et~al.}(2015)\citenamefont {Fuchs},
  \citenamefont {Stender}, \citenamefont {Trupke}, \citenamefont {Simin},
  \citenamefont {Pflaum}, \citenamefont {Dyakonov},\ and\ \citenamefont
  {Astakhov}}]{Fuchs2015}%
  \BibitemOpen
  \bibfield  {author} {\bibinfo {author} {\bibfnamefont {F.}~\bibnamefont
  {Fuchs}}, \bibinfo {author} {\bibfnamefont {B.}~\bibnamefont {Stender}},
  \bibinfo {author} {\bibfnamefont {M.}~\bibnamefont {Trupke}}, \bibinfo
  {author} {\bibfnamefont {D.}~\bibnamefont {Simin}}, \bibinfo {author}
  {\bibfnamefont {J.}~\bibnamefont {Pflaum}}, \bibinfo {author} {\bibfnamefont
  {V.}~\bibnamefont {Dyakonov}},\ and\ \bibinfo {author} {\bibfnamefont
  {G.~V.}\ \bibnamefont {Astakhov}},\ }\bibfield  {title} {\bibinfo {title}
  {{Engineering near-infrared single-photon emitters with optically active
  spins in ultrapure silicon carbide}},\ }\href
  {https://doi.org/10.1038/ncomms8578} {\bibfield  {journal} {\bibinfo
  {journal} {Nature Communications}\ }\textbf {\bibinfo {volume} {6}},\
  \bibinfo {pages} {7578} (\bibinfo {year} {2015})}\BibitemShut {NoStop}%
\bibitem [{\citenamefont {Capelli}\ \emph {et~al.}(2019)\citenamefont
  {Capelli}, \citenamefont {Heffernan}, \citenamefont {Ohshima}, \citenamefont
  {Abe}, \citenamefont {Jeske}, \citenamefont {Hope}, \citenamefont
  {Greentree}, \citenamefont {Reineck},\ and\ \citenamefont
  {Gibson}}]{Capelli2019}%
  \BibitemOpen
  \bibfield  {author} {\bibinfo {author} {\bibfnamefont {M.}~\bibnamefont
  {Capelli}}, \bibinfo {author} {\bibfnamefont {A.~H.}\ \bibnamefont
  {Heffernan}}, \bibinfo {author} {\bibfnamefont {T.}~\bibnamefont {Ohshima}},
  \bibinfo {author} {\bibfnamefont {H.}~\bibnamefont {Abe}}, \bibinfo {author}
  {\bibfnamefont {J.}~\bibnamefont {Jeske}}, \bibinfo {author} {\bibfnamefont
  {A.}~\bibnamefont {Hope}}, \bibinfo {author} {\bibfnamefont {A.~D.}\
  \bibnamefont {Greentree}}, \bibinfo {author} {\bibfnamefont {P.}~\bibnamefont
  {Reineck}},\ and\ \bibinfo {author} {\bibfnamefont {B.~C.}\ \bibnamefont
  {Gibson}},\ }\bibfield  {title} {\bibinfo {title} {{Increased
  nitrogen-vacancy centre creation yield in diamond through electron beam
  irradiation at high temperature}},\ }\href
  {https://doi.org/10.1016/j.carbon.2018.11.051} {\bibfield  {journal}
  {\bibinfo  {journal} {Carbon}\ }\textbf {\bibinfo {volume} {143}},\ \bibinfo
  {pages} {714} (\bibinfo {year} {2019})}\BibitemShut {NoStop}%
\bibitem [{\citenamefont {Xu}\ \emph {et~al.}(2018)\citenamefont {Xu},
  \citenamefont {Elbadawi}, \citenamefont {Tran}, \citenamefont {Kianinia},
  \citenamefont {Li}, \citenamefont {Liu}, \citenamefont {Hoffman},
  \citenamefont {Nguyen}, \citenamefont {Kim}, \citenamefont {Edgar},
  \citenamefont {Wu}, \citenamefont {Song}, \citenamefont {Ali}, \citenamefont
  {Ford}, \citenamefont {Toth},\ and\ \citenamefont {Aharonovich}}]{Xu2018}%
  \BibitemOpen
  \bibfield  {author} {\bibinfo {author} {\bibfnamefont {Z.~Q.}\ \bibnamefont
  {Xu}}, \bibinfo {author} {\bibfnamefont {C.}~\bibnamefont {Elbadawi}},
  \bibinfo {author} {\bibfnamefont {T.~T.}\ \bibnamefont {Tran}}, \bibinfo
  {author} {\bibfnamefont {M.}~\bibnamefont {Kianinia}}, \bibinfo {author}
  {\bibfnamefont {X.}~\bibnamefont {Li}}, \bibinfo {author} {\bibfnamefont
  {D.}~\bibnamefont {Liu}}, \bibinfo {author} {\bibfnamefont {T.~B.}\
  \bibnamefont {Hoffman}}, \bibinfo {author} {\bibfnamefont {M.}~\bibnamefont
  {Nguyen}}, \bibinfo {author} {\bibfnamefont {S.}~\bibnamefont {Kim}},
  \bibinfo {author} {\bibfnamefont {J.~H.}\ \bibnamefont {Edgar}}, \bibinfo
  {author} {\bibfnamefont {X.}~\bibnamefont {Wu}}, \bibinfo {author}
  {\bibfnamefont {L.}~\bibnamefont {Song}}, \bibinfo {author} {\bibfnamefont
  {S.}~\bibnamefont {Ali}}, \bibinfo {author} {\bibfnamefont {M.}~\bibnamefont
  {Ford}}, \bibinfo {author} {\bibfnamefont {M.}~\bibnamefont {Toth}},\ and\
  \bibinfo {author} {\bibfnamefont {I.}~\bibnamefont {Aharonovich}},\
  }\bibfield  {title} {\bibinfo {title} {{Single photon emission from plasma
  treated 2D hexagonal boron nitride}},\ }\href
  {https://doi.org/10.1039/c7nr08222c} {\bibfield  {journal} {\bibinfo
  {journal} {Nanoscale}\ }\textbf {\bibinfo {volume} {10}},\ \bibinfo {pages}
  {7957} (\bibinfo {year} {2018})}\BibitemShut {NoStop}%
\bibitem [{\citenamefont {Lyu}\ \emph {et~al.}(2020)\citenamefont {Lyu},
  \citenamefont {Zhu}, \citenamefont {Gu}, \citenamefont {Qiao}, \citenamefont
  {Watanabe}, \citenamefont {Taniguchi},\ and\ \citenamefont {Ye}}]{Lyu2020}%
  \BibitemOpen
  \bibfield  {author} {\bibinfo {author} {\bibfnamefont {C.}~\bibnamefont
  {Lyu}}, \bibinfo {author} {\bibfnamefont {Y.}~\bibnamefont {Zhu}}, \bibinfo
  {author} {\bibfnamefont {P.}~\bibnamefont {Gu}}, \bibinfo {author}
  {\bibfnamefont {J.}~\bibnamefont {Qiao}}, \bibinfo {author} {\bibfnamefont
  {K.}~\bibnamefont {Watanabe}}, \bibinfo {author} {\bibfnamefont
  {T.}~\bibnamefont {Taniguchi}},\ and\ \bibinfo {author} {\bibfnamefont
  {Y.}~\bibnamefont {Ye}},\ }\bibfield  {title} {\bibinfo {title}
  {{Single-photon emission from two-dimensional hexagonal boron nitride
  annealed in a carbon-rich environment}},\ }\href
  {https://doi.org/10.1063/5.0025792} {\bibfield  {journal} {\bibinfo
  {journal} {Applied Physics Letters}\ }\textbf {\bibinfo {volume} {117}},\
  \bibinfo {pages} {244002} (\bibinfo {year} {2020})}\BibitemShut {NoStop}%
\bibitem [{\citenamefont {Kim}\ \emph {et~al.}(2019)\citenamefont {Kim},
  \citenamefont {Moon}, \citenamefont {Noh}, \citenamefont {Lee},\ and\
  \citenamefont {Kim}}]{Kim2019}%
  \BibitemOpen
  \bibfield  {author} {\bibinfo {author} {\bibfnamefont {H.}~\bibnamefont
  {Kim}}, \bibinfo {author} {\bibfnamefont {J.~S.}\ \bibnamefont {Moon}},
  \bibinfo {author} {\bibfnamefont {G.}~\bibnamefont {Noh}}, \bibinfo {author}
  {\bibfnamefont {J.}~\bibnamefont {Lee}},\ and\ \bibinfo {author}
  {\bibfnamefont {J.~H.}\ \bibnamefont {Kim}},\ }\bibfield  {title} {\bibinfo
  {title} {{Position and Frequency Control of Strain-Induced Quantum Emitters
  in WSe2 Monolayers}},\ }\href {https://doi.org/10.1021/acs.nanolett.9b03421}
  {\bibfield  {journal} {\bibinfo  {journal} {Nano Letters}\ }\textbf {\bibinfo
  {volume} {19}},\ \bibinfo {pages} {7534} (\bibinfo {year}
  {2019})}\BibitemShut {NoStop}%
\bibitem [{\citenamefont {Schell}\ \emph {et~al.}(2016)\citenamefont {Schell},
  \citenamefont {Tran}, \citenamefont {Takashima}, \citenamefont {Takeuchi},\
  and\ \citenamefont {Aharonovich}}]{Schell2016}%
  \BibitemOpen
  \bibfield  {author} {\bibinfo {author} {\bibfnamefont {A.~W.}\ \bibnamefont
  {Schell}}, \bibinfo {author} {\bibfnamefont {T.~T.}\ \bibnamefont {Tran}},
  \bibinfo {author} {\bibfnamefont {H.}~\bibnamefont {Takashima}}, \bibinfo
  {author} {\bibfnamefont {S.}~\bibnamefont {Takeuchi}},\ and\ \bibinfo
  {author} {\bibfnamefont {I.}~\bibnamefont {Aharonovich}},\ }\bibfield
  {title} {\bibinfo {title} {{Non-linear excitation of quantum emitters in
  hexagonal boron nitride multiplayers}},\ }\href
  {https://doi.org/10.1063/1.4961684} {\bibfield  {journal} {\bibinfo
  {journal} {APL Photonics}\ }\textbf {\bibinfo {volume} {1}},\ \bibinfo
  {pages} {091302} (\bibinfo {year} {2016})}\BibitemShut {NoStop}%
\bibitem [{\citenamefont {Breitweiser}\ \emph {et~al.}(2020)\citenamefont
  {Breitweiser}, \citenamefont {Exarhos}, \citenamefont {Patel}, \citenamefont
  {Saouaf}, \citenamefont {Porat}, \citenamefont {Hopper},\ and\ \citenamefont
  {Bassett}}]{Breitweiser2020}%
  \BibitemOpen
  \bibfield  {author} {\bibinfo {author} {\bibfnamefont {S.~A.}\ \bibnamefont
  {Breitweiser}}, \bibinfo {author} {\bibfnamefont {A.~L.}\ \bibnamefont
  {Exarhos}}, \bibinfo {author} {\bibfnamefont {R.~N.}\ \bibnamefont {Patel}},
  \bibinfo {author} {\bibfnamefont {J.}~\bibnamefont {Saouaf}}, \bibinfo
  {author} {\bibfnamefont {B.}~\bibnamefont {Porat}}, \bibinfo {author}
  {\bibfnamefont {D.~A.}\ \bibnamefont {Hopper}},\ and\ \bibinfo {author}
  {\bibfnamefont {L.~C.}\ \bibnamefont {Bassett}},\ }\bibfield  {title}
  {\bibinfo {title} {{Efficient Optical Quantification of Heterogeneous Emitter
  Ensembles}},\ }\href {https://doi.org/10.1021/acsphotonics.9b01707}
  {\bibfield  {journal} {\bibinfo  {journal} {ACS Photonics}\ }\textbf
  {\bibinfo {volume} {7}},\ \bibinfo {pages} {288} (\bibinfo {year}
  {2020})}\BibitemShut {NoStop}%
\bibitem [{\citenamefont {Dodson}\ \emph {et~al.}(2014)\citenamefont {Dodson},
  \citenamefont {Kurvits}, \citenamefont {Li},\ and\ \citenamefont
  {Zia}}]{Dodson2014}%
  \BibitemOpen
  \bibfield  {author} {\bibinfo {author} {\bibfnamefont {C.~M.}\ \bibnamefont
  {Dodson}}, \bibinfo {author} {\bibfnamefont {J.~A.}\ \bibnamefont {Kurvits}},
  \bibinfo {author} {\bibfnamefont {D.}~\bibnamefont {Li}},\ and\ \bibinfo
  {author} {\bibfnamefont {R.}~\bibnamefont {Zia}},\ }\bibfield  {title}
  {\bibinfo {title} {{Wide-angle energy-momentum spectroscopy}},\ }\href
  {https://doi.org/10.1364/ol.39.003927} {\bibfield  {journal} {\bibinfo
  {journal} {Optics Letters}\ }\textbf {\bibinfo {volume} {39}},\ \bibinfo
  {pages} {3927} (\bibinfo {year} {2014})}\BibitemShut {NoStop}%
\bibitem [{\citenamefont {Fishman}\ \emph {et~al.}(2021)\citenamefont
  {Fishman}, \citenamefont {Patel}, \citenamefont {Hopper}, \citenamefont
  {Huang},\ and\ \citenamefont {Bassett}}]{fishman2021photon}%
  \BibitemOpen
  \bibfield  {author} {\bibinfo {author} {\bibfnamefont {R.~E.~K.}\
  \bibnamefont {Fishman}}, \bibinfo {author} {\bibfnamefont {R.~N.}\
  \bibnamefont {Patel}}, \bibinfo {author} {\bibfnamefont {D.~A.}\ \bibnamefont
  {Hopper}}, \bibinfo {author} {\bibfnamefont {T.-Y.}\ \bibnamefont {Huang}},\
  and\ \bibinfo {author} {\bibfnamefont {L.~C.}\ \bibnamefont {Bassett}},\
  }\bibfield  {title} {\bibinfo {title} {Photon emission correlation
  spectroscopy as an analytical tool for quantum defects},\ }\href@noop {}
  {\bibfield  {journal} {\bibinfo  {journal} {arXiv preprint arXiv:2111.01252}\
  } (\bibinfo {year} {2021})},\ \Eprint
  {https://arxiv.org/abs/https://arxiv.org/abs/2111.01252}
  {https://arxiv.org/abs/2111.01252} \BibitemShut {NoStop}%
\bibitem [{\citenamefont {Kudyshev}\ \emph {et~al.}(2020)\citenamefont
  {Kudyshev}, \citenamefont {Bogdanov}, \citenamefont {Isacsson}, \citenamefont
  {Kildishev}, \citenamefont {Boltasseva},\ and\ \citenamefont
  {Shalaev}}]{Kudyshev2020}%
  \BibitemOpen
  \bibfield  {author} {\bibinfo {author} {\bibfnamefont {Z.~A.}\ \bibnamefont
  {Kudyshev}}, \bibinfo {author} {\bibfnamefont {S.~I.}\ \bibnamefont
  {Bogdanov}}, \bibinfo {author} {\bibfnamefont {T.}~\bibnamefont {Isacsson}},
  \bibinfo {author} {\bibfnamefont {A.~V.}\ \bibnamefont {Kildishev}}, \bibinfo
  {author} {\bibfnamefont {A.}~\bibnamefont {Boltasseva}},\ and\ \bibinfo
  {author} {\bibfnamefont {V.~M.}\ \bibnamefont {Shalaev}},\ }\bibfield
  {title} {\bibinfo {title} {{Rapid Classification of Quantum Sources Enabled
  by Machine Learning}},\ }\href {https://doi.org/10.1002/qute.202000067}
  {\bibfield  {journal} {\bibinfo  {journal} {Advanced Quantum Technologies}\
  }\textbf {\bibinfo {volume} {3}},\ \bibinfo {pages} {2000067} (\bibinfo
  {year} {2020})}\BibitemShut {NoStop}%
\bibitem [{\citenamefont {{Van Dam}}\ \emph {et~al.}(2019)\citenamefont {{Van
  Dam}}, \citenamefont {Walsh}, \citenamefont {Degen}, \citenamefont {Bersin},
  \citenamefont {Mouradian}, \citenamefont {Galiullin}, \citenamefont {Ruf},
  \citenamefont {Ijspeert}, \citenamefont {Taminiau}, \citenamefont {Hanson},\
  and\ \citenamefont {Englund}}]{VanDam2019}%
  \BibitemOpen
  \bibfield  {author} {\bibinfo {author} {\bibfnamefont {S.~B.}\ \bibnamefont
  {{Van Dam}}}, \bibinfo {author} {\bibfnamefont {M.}~\bibnamefont {Walsh}},
  \bibinfo {author} {\bibfnamefont {M.~J.}\ \bibnamefont {Degen}}, \bibinfo
  {author} {\bibfnamefont {E.}~\bibnamefont {Bersin}}, \bibinfo {author}
  {\bibfnamefont {S.~L.}\ \bibnamefont {Mouradian}}, \bibinfo {author}
  {\bibfnamefont {A.}~\bibnamefont {Galiullin}}, \bibinfo {author}
  {\bibfnamefont {M.}~\bibnamefont {Ruf}}, \bibinfo {author} {\bibfnamefont
  {M.}~\bibnamefont {Ijspeert}}, \bibinfo {author} {\bibfnamefont {T.~H.}\
  \bibnamefont {Taminiau}}, \bibinfo {author} {\bibfnamefont {R.}~\bibnamefont
  {Hanson}},\ and\ \bibinfo {author} {\bibfnamefont {D.~R.}\ \bibnamefont
  {Englund}},\ }\bibfield  {title} {\bibinfo {title} {{Optical coherence of
  diamond nitrogen-vacancy centers formed by ion implantation and annealing}},\
  }\href {https://doi.org/10.1103/PhysRevB.99.161203} {\bibfield  {journal}
  {\bibinfo  {journal} {Physical Review B}\ }\textbf {\bibinfo {volume} {99}},\
  \bibinfo {pages} {161203} (\bibinfo {year} {2019})}\BibitemShut {NoStop}%
\bibitem [{\citenamefont {Chen}\ \emph {et~al.}(2013)\citenamefont {Chen},
  \citenamefont {Liu}, \citenamefont {Song}, \citenamefont {Wu}, \citenamefont
  {Wu},\ and\ \citenamefont {Zeng}}]{Chen2013}%
  \BibitemOpen
  \bibfield  {author} {\bibinfo {author} {\bibfnamefont {G.}~\bibnamefont
  {Chen}}, \bibinfo {author} {\bibfnamefont {Y.}~\bibnamefont {Liu}}, \bibinfo
  {author} {\bibfnamefont {M.}~\bibnamefont {Song}}, \bibinfo {author}
  {\bibfnamefont {B.}~\bibnamefont {Wu}}, \bibinfo {author} {\bibfnamefont
  {E.}~\bibnamefont {Wu}},\ and\ \bibinfo {author} {\bibfnamefont
  {H.}~\bibnamefont {Zeng}},\ }\bibfield  {title} {\bibinfo {title}
  {{Photoluminescence enhancement dependent on the orientations of single NV
  centers in nanodiamonds on a gold film}},\ }\href
  {https://doi.org/10.1109/JSTQE.2013.2247977} {\bibfield  {journal} {\bibinfo
  {journal} {IEEE J. Sel. Top. Quantum Electron.}\ }\textbf {\bibinfo {volume}
  {19}},\ \bibinfo {pages} {4602404} (\bibinfo {year} {2013})}\BibitemShut
  {NoStop}%
\bibitem [{\citenamefont {Cai}\ \emph {et~al.}(2018)\citenamefont {Cai},
  \citenamefont {Kim}, \citenamefont {Yang}, \citenamefont {Dutta},
  \citenamefont {Aghaeimeibodi},\ and\ \citenamefont {Waks}}]{Cai2018a}%
  \BibitemOpen
  \bibfield  {author} {\bibinfo {author} {\bibfnamefont {T.}~\bibnamefont
  {Cai}}, \bibinfo {author} {\bibfnamefont {J.~H.}\ \bibnamefont {Kim}},
  \bibinfo {author} {\bibfnamefont {Z.}~\bibnamefont {Yang}}, \bibinfo {author}
  {\bibfnamefont {S.}~\bibnamefont {Dutta}}, \bibinfo {author} {\bibfnamefont
  {S.}~\bibnamefont {Aghaeimeibodi}},\ and\ \bibinfo {author} {\bibfnamefont
  {E.}~\bibnamefont {Waks}},\ }\bibfield  {title} {\bibinfo {title} {{Radiative
  Enhancement of Single Quantum Emitters in WSe2 Monolayers Using
  Site-Controlled Metallic Nanopillars}},\ }\href
  {https://doi.org/10.1021/acsphotonics.8b00580} {\bibfield  {journal}
  {\bibinfo  {journal} {ACS Photonics}\ }\textbf {\bibinfo {volume} {5}},\
  \bibinfo {pages} {3466} (\bibinfo {year} {2018})}\BibitemShut {NoStop}%
\bibitem [{\citenamefont {Liu}\ \emph {et~al.}(2015)\citenamefont {Liu},
  \citenamefont {Chen}, \citenamefont {Rong}, \citenamefont {McGuinness},
  \citenamefont {Jelezko}, \citenamefont {Tamura}, \citenamefont {Tanii},
  \citenamefont {Teraji}, \citenamefont {Onoda}, \citenamefont {Ohshima},
  \citenamefont {Isoya}, \citenamefont {Shinada}, \citenamefont {Wu},\ and\
  \citenamefont {Zeng}}]{Liu2015}%
  \BibitemOpen
  \bibfield  {author} {\bibinfo {author} {\bibfnamefont {Y.}~\bibnamefont
  {Liu}}, \bibinfo {author} {\bibfnamefont {G.}~\bibnamefont {Chen}}, \bibinfo
  {author} {\bibfnamefont {Y.}~\bibnamefont {Rong}}, \bibinfo {author}
  {\bibfnamefont {L.~P.}\ \bibnamefont {McGuinness}}, \bibinfo {author}
  {\bibfnamefont {F.}~\bibnamefont {Jelezko}}, \bibinfo {author} {\bibfnamefont
  {S.}~\bibnamefont {Tamura}}, \bibinfo {author} {\bibfnamefont
  {T.}~\bibnamefont {Tanii}}, \bibinfo {author} {\bibfnamefont
  {T.}~\bibnamefont {Teraji}}, \bibinfo {author} {\bibfnamefont
  {S.}~\bibnamefont {Onoda}}, \bibinfo {author} {\bibfnamefont
  {T.}~\bibnamefont {Ohshima}}, \bibinfo {author} {\bibfnamefont
  {J.}~\bibnamefont {Isoya}}, \bibinfo {author} {\bibfnamefont
  {T.}~\bibnamefont {Shinada}}, \bibinfo {author} {\bibfnamefont
  {E.}~\bibnamefont {Wu}},\ and\ \bibinfo {author} {\bibfnamefont
  {H.}~\bibnamefont {Zeng}},\ }\bibfield  {title} {\bibinfo {title}
  {{Fluorescence Polarization Switching from a Single Silicon Vacancy Colour
  Centre in Diamond}},\ }\href {https://doi.org/10.1038/srep12244} {\bibfield
  {journal} {\bibinfo  {journal} {Scientific Reports}\ }\textbf {\bibinfo
  {volume} {5}},\ \bibinfo {pages} {1} (\bibinfo {year} {2015})}\BibitemShut
  {NoStop}%
\bibitem [{\citenamefont {{Gatto Monticone}}\ \emph {et~al.}(2014)\citenamefont
  {{Gatto Monticone}}, \citenamefont {Traina}, \citenamefont {Moreva},
  \citenamefont {Forneris}, \citenamefont {Olivero}, \citenamefont
  {Degiovanni}, \citenamefont {Taccetti}, \citenamefont {Giuntini},
  \citenamefont {Brida}, \citenamefont {Amato},\ and\ \citenamefont
  {Genovese}}]{GattoMonticone2014}%
  \BibitemOpen
  \bibfield  {author} {\bibinfo {author} {\bibfnamefont {D.}~\bibnamefont
  {{Gatto Monticone}}}, \bibinfo {author} {\bibfnamefont {P.}~\bibnamefont
  {Traina}}, \bibinfo {author} {\bibfnamefont {E.}~\bibnamefont {Moreva}},
  \bibinfo {author} {\bibfnamefont {J.}~\bibnamefont {Forneris}}, \bibinfo
  {author} {\bibfnamefont {P.}~\bibnamefont {Olivero}}, \bibinfo {author}
  {\bibfnamefont {I.~P.}\ \bibnamefont {Degiovanni}}, \bibinfo {author}
  {\bibfnamefont {F.}~\bibnamefont {Taccetti}}, \bibinfo {author}
  {\bibfnamefont {L.}~\bibnamefont {Giuntini}}, \bibinfo {author}
  {\bibfnamefont {G.}~\bibnamefont {Brida}}, \bibinfo {author} {\bibfnamefont
  {G.}~\bibnamefont {Amato}},\ and\ \bibinfo {author} {\bibfnamefont
  {M.}~\bibnamefont {Genovese}},\ }\bibfield  {title} {\bibinfo {title}
  {{Native NIR-emitting single colour centres in CVD diamond}},\ }\href
  {https://doi.org/10.1088/1367-2630/16/5/053005} {\bibfield  {journal}
  {\bibinfo  {journal} {New Journal of Physics}\ }\textbf {\bibinfo {volume}
  {16}},\ \bibinfo {pages} {053005} (\bibinfo {year} {2014})}\BibitemShut
  {NoStop}%
\bibitem [{\citenamefont {Choi}\ \emph {et~al.}(2016)\citenamefont {Choi},
  \citenamefont {Tran}, \citenamefont {Elbadawi}, \citenamefont {Lobo},
  \citenamefont {Wang}, \citenamefont {Juodkazis}, \citenamefont {Seniutinas},
  \citenamefont {Toth},\ and\ \citenamefont {Aharonovich}}]{Choi2016}%
  \BibitemOpen
  \bibfield  {author} {\bibinfo {author} {\bibfnamefont {S.}~\bibnamefont
  {Choi}}, \bibinfo {author} {\bibfnamefont {T.~T.}\ \bibnamefont {Tran}},
  \bibinfo {author} {\bibfnamefont {C.}~\bibnamefont {Elbadawi}}, \bibinfo
  {author} {\bibfnamefont {C.}~\bibnamefont {Lobo}}, \bibinfo {author}
  {\bibfnamefont {X.}~\bibnamefont {Wang}}, \bibinfo {author} {\bibfnamefont
  {S.}~\bibnamefont {Juodkazis}}, \bibinfo {author} {\bibfnamefont
  {G.}~\bibnamefont {Seniutinas}}, \bibinfo {author} {\bibfnamefont
  {M.}~\bibnamefont {Toth}},\ and\ \bibinfo {author} {\bibfnamefont
  {I.}~\bibnamefont {Aharonovich}},\ }\bibfield  {title} {\bibinfo {title}
  {{Engineering and Localization of Quantum Emitters in Large Hexagonal Boron
  Nitride Layers}},\ }\href {https://doi.org/10.1021/acsami.6b09875} {\bibfield
   {journal} {\bibinfo  {journal} {ACS Applied Materials and Interfaces}\
  }\textbf {\bibinfo {volume} {8}},\ \bibinfo {pages} {29642} (\bibinfo {year}
  {2016})}\BibitemShut {NoStop}%
\bibitem [{\citenamefont {Himics}\ \emph {et~al.}(2014)\citenamefont {Himics},
  \citenamefont {T{\'{o}}th}, \citenamefont {Veres}, \citenamefont {Balogh},\
  and\ \citenamefont {Ko{\'{o}}s}}]{Himics2014}%
  \BibitemOpen
  \bibfield  {author} {\bibinfo {author} {\bibfnamefont {L.}~\bibnamefont
  {Himics}}, \bibinfo {author} {\bibfnamefont {S.}~\bibnamefont {T{\'{o}}th}},
  \bibinfo {author} {\bibfnamefont {M.}~\bibnamefont {Veres}}, \bibinfo
  {author} {\bibfnamefont {Z.}~\bibnamefont {Balogh}},\ and\ \bibinfo {author}
  {\bibfnamefont {M.}~\bibnamefont {Ko{\'{o}}s}},\ }\bibfield  {title}
  {\bibinfo {title} {{Creation of deep blue light emitting nitrogen-vacancy
  center in nanosized diamond}},\ }\href {https://doi.org/10.1063/1.4867463}
  {\bibfield  {journal} {\bibinfo  {journal} {Applied Physics Letters}\
  }\textbf {\bibinfo {volume} {104}},\ \bibinfo {pages} {093101} (\bibinfo
  {year} {2014})}\BibitemShut {NoStop}%
\bibitem [{\citenamefont {Schr{\"{o}}der}\ \emph {et~al.}(2017)\citenamefont
  {Schr{\"{o}}der}, \citenamefont {Trusheim}, \citenamefont {Walsh},
  \citenamefont {Li}, \citenamefont {Zheng}, \citenamefont {Schukraft},
  \citenamefont {Sipahigil}, \citenamefont {Evans}, \citenamefont {Sukachev},
  \citenamefont {Nguyen}, \citenamefont {Pacheco}, \citenamefont {Camacho},
  \citenamefont {Bielejec}, \citenamefont {Lukin},\ and\ \citenamefont
  {Englund}}]{Schroder2017}%
  \BibitemOpen
  \bibfield  {author} {\bibinfo {author} {\bibfnamefont {T.}~\bibnamefont
  {Schr{\"{o}}der}}, \bibinfo {author} {\bibfnamefont {M.~E.}\ \bibnamefont
  {Trusheim}}, \bibinfo {author} {\bibfnamefont {M.}~\bibnamefont {Walsh}},
  \bibinfo {author} {\bibfnamefont {L.}~\bibnamefont {Li}}, \bibinfo {author}
  {\bibfnamefont {J.}~\bibnamefont {Zheng}}, \bibinfo {author} {\bibfnamefont
  {M.}~\bibnamefont {Schukraft}}, \bibinfo {author} {\bibfnamefont
  {A.}~\bibnamefont {Sipahigil}}, \bibinfo {author} {\bibfnamefont {R.~E.}\
  \bibnamefont {Evans}}, \bibinfo {author} {\bibfnamefont {D.~D.}\ \bibnamefont
  {Sukachev}}, \bibinfo {author} {\bibfnamefont {C.~T.}\ \bibnamefont
  {Nguyen}}, \bibinfo {author} {\bibfnamefont {J.~L.}\ \bibnamefont {Pacheco}},
  \bibinfo {author} {\bibfnamefont {R.~M.}\ \bibnamefont {Camacho}}, \bibinfo
  {author} {\bibfnamefont {E.~S.}\ \bibnamefont {Bielejec}}, \bibinfo {author}
  {\bibfnamefont {M.~D.}\ \bibnamefont {Lukin}},\ and\ \bibinfo {author}
  {\bibfnamefont {D.}~\bibnamefont {Englund}},\ }\bibfield  {title} {\bibinfo
  {title} {{Scalable focused ion beam creation of nearly lifetime-limited
  single quantum emitters in diamond nanostructures}},\ }\href
  {https://doi.org/10.1038/ncomms15376} {\bibfield  {journal} {\bibinfo
  {journal} {Nature Communications}\ }\textbf {\bibinfo {volume} {8}},\
  \bibinfo {pages} {1} (\bibinfo {year} {2017})}\BibitemShut {NoStop}%
\bibitem [{\citenamefont {Thiruraman}\ \emph {et~al.}(2019)\citenamefont
  {Thiruraman}, \citenamefont {{Masih Das}},\ and\ \citenamefont
  {Drndi{\'{c}}}}]{Thiruraman2019}%
  \BibitemOpen
  \bibfield  {author} {\bibinfo {author} {\bibfnamefont {J.~P.}\ \bibnamefont
  {Thiruraman}}, \bibinfo {author} {\bibfnamefont {P.}~\bibnamefont {{Masih
  Das}}},\ and\ \bibinfo {author} {\bibfnamefont {M.}~\bibnamefont
  {Drndi{\'{c}}}},\ }\bibfield  {title} {\bibinfo {title} {{Irradiation of
  Transition Metal Dichalcogenides Using a Focused Ion Beam: Controlled
  Single-Atom Defect Creation}},\ }\href
  {https://doi.org/10.1002/adfm.201904668} {\bibfield  {journal} {\bibinfo
  {journal} {Advanced Functional Materials}\ }\textbf {\bibinfo {volume}
  {29}},\ \bibinfo {pages} {1904668} (\bibinfo {year} {2019})}\BibitemShut
  {NoStop}%
\bibitem [{\citenamefont {Shulevitz}\ \emph {et~al.}(2021)\citenamefont
  {Shulevitz}, \citenamefont {Huang}, \citenamefont {Xu}, \citenamefont
  {Neuhaus}, \citenamefont {Patel}, \citenamefont {Bassett},\ and\
  \citenamefont {Kagan}}]{shulevitz2021template}%
  \BibitemOpen
  \bibfield  {author} {\bibinfo {author} {\bibfnamefont {H.~J.}\ \bibnamefont
  {Shulevitz}}, \bibinfo {author} {\bibfnamefont {T.-Y.}\ \bibnamefont
  {Huang}}, \bibinfo {author} {\bibfnamefont {J.}~\bibnamefont {Xu}}, \bibinfo
  {author} {\bibfnamefont {S.}~\bibnamefont {Neuhaus}}, \bibinfo {author}
  {\bibfnamefont {R.~N.}\ \bibnamefont {Patel}}, \bibinfo {author}
  {\bibfnamefont {L.~C.}\ \bibnamefont {Bassett}},\ and\ \bibinfo {author}
  {\bibfnamefont {C.~R.}\ \bibnamefont {Kagan}},\ }\bibfield  {title} {\bibinfo
  {title} {Template-assisted self assembly of fluorescent nanodiamonds for
  scalable quantum technologies},\ }\href@noop {} {\bibfield  {journal}
  {\bibinfo  {journal} {arXiv preprint arXiv:2111.14921}\ } (\bibinfo {year}
  {2021})},\ \Eprint {https://arxiv.org/abs/https://arxiv.org/abs/2111.14921}
  {https://arxiv.org/abs/2111.14921} \BibitemShut {NoStop}%
\bibitem [{\citenamefont {Manzo}\ and\ \citenamefont
  {Garcia-Parajo}(2015)}]{Manzo2015}%
  \BibitemOpen
  \bibfield  {author} {\bibinfo {author} {\bibfnamefont {C.}~\bibnamefont
  {Manzo}}\ and\ \bibinfo {author} {\bibfnamefont {M.~F.}\ \bibnamefont
  {Garcia-Parajo}},\ }\bibfield  {title} {\bibinfo {title} {{A review of
  progress in single particle tracking: From methods to biophysical
  insights}},\ }\href {https://doi.org/10.1088/0034-4885/78/12/124601}
  {\bibfield  {journal} {\bibinfo  {journal} {Reports on Progress in Physics}\
  }\textbf {\bibinfo {volume} {78}},\ \bibinfo {pages} {124601} (\bibinfo
  {year} {2015})}\BibitemShut {NoStop}%
\bibitem [{\citenamefont {Schermelleh}\ \emph {et~al.}(2019)\citenamefont
  {Schermelleh}, \citenamefont {Ferrand}, \citenamefont {Huser}, \citenamefont
  {Eggeling}, \citenamefont {Sauer}, \citenamefont {Biehlmaier},\ and\
  \citenamefont {Drummen}}]{Schermelleh2019}%
  \BibitemOpen
  \bibfield  {author} {\bibinfo {author} {\bibfnamefont {L.}~\bibnamefont
  {Schermelleh}}, \bibinfo {author} {\bibfnamefont {A.}~\bibnamefont
  {Ferrand}}, \bibinfo {author} {\bibfnamefont {T.}~\bibnamefont {Huser}},
  \bibinfo {author} {\bibfnamefont {C.}~\bibnamefont {Eggeling}}, \bibinfo
  {author} {\bibfnamefont {M.}~\bibnamefont {Sauer}}, \bibinfo {author}
  {\bibfnamefont {O.}~\bibnamefont {Biehlmaier}},\ and\ \bibinfo {author}
  {\bibfnamefont {G.~P.}\ \bibnamefont {Drummen}},\ }\bibfield  {title}
  {\bibinfo {title} {{Super-resolution microscopy demystified}},\ }\href
  {https://doi.org/10.1038/s41556-018-0251-8} {\bibfield  {journal} {\bibinfo
  {journal} {Nature Cell Biology}\ }\textbf {\bibinfo {volume} {21}},\ \bibinfo
  {pages} {72} (\bibinfo {year} {2019})}\BibitemShut {NoStop}%
\bibitem [{\citenamefont {Bradley}\ and\ \citenamefont
  {Roth}(2007)}]{Bradley2007}%
  \BibitemOpen
  \bibfield  {author} {\bibinfo {author} {\bibfnamefont {D.}~\bibnamefont
  {Bradley}}\ and\ \bibinfo {author} {\bibfnamefont {G.}~\bibnamefont {Roth}},\
  }\bibfield  {title} {\bibinfo {title} {{Adaptive Thresholding using the
  Integral Image}},\ }\href {https://doi.org/10.1080/2151237x.2007.10129236}
  {\bibfield  {journal} {\bibinfo  {journal} {Journal of Graphics Tools}\
  }\textbf {\bibinfo {volume} {12}},\ \bibinfo {pages} {13} (\bibinfo {year}
  {2007})}\BibitemShut {NoStop}%
\bibitem [{\citenamefont {Zhang}\ \emph {et~al.}(2007)\citenamefont {Zhang},
  \citenamefont {Zerubia},\ and\ \citenamefont {Olivo-Marin}}]{Zhang2007}%
  \BibitemOpen
  \bibfield  {author} {\bibinfo {author} {\bibfnamefont {B.}~\bibnamefont
  {Zhang}}, \bibinfo {author} {\bibfnamefont {J.}~\bibnamefont {Zerubia}},\
  and\ \bibinfo {author} {\bibfnamefont {J.~C.}\ \bibnamefont {Olivo-Marin}},\
  }\bibfield  {title} {\bibinfo {title} {{Gaussian approximations of
  fluorescence microscope point-spread function models}},\ }\href
  {https://doi.org/10.1364/AO.46.001819} {\bibfield  {journal} {\bibinfo
  {journal} {Applied Optics}\ }\textbf {\bibinfo {volume} {46}},\ \bibinfo
  {pages} {1819} (\bibinfo {year} {2007})}\BibitemShut {NoStop}%
\bibitem [{\citenamefont {Fukushige}\ \emph {et~al.}(2020)\citenamefont
  {Fukushige}, \citenamefont {Kawaguchi}, \citenamefont {Shimazaki},
  \citenamefont {Tashima}, \citenamefont {Takashima},\ and\ \citenamefont
  {Takeuchi}}]{Fukushige2020}%
  \BibitemOpen
  \bibfield  {author} {\bibinfo {author} {\bibfnamefont {K.}~\bibnamefont
  {Fukushige}}, \bibinfo {author} {\bibfnamefont {H.}~\bibnamefont
  {Kawaguchi}}, \bibinfo {author} {\bibfnamefont {K.}~\bibnamefont
  {Shimazaki}}, \bibinfo {author} {\bibfnamefont {T.}~\bibnamefont {Tashima}},
  \bibinfo {author} {\bibfnamefont {H.}~\bibnamefont {Takashima}},\ and\
  \bibinfo {author} {\bibfnamefont {S.}~\bibnamefont {Takeuchi}},\ }\bibfield
  {title} {\bibinfo {title} {{Identification of the orientation of a single NV
  center in a nanodiamond using a three-dimensionally controlled magnetic
  field}},\ }\href {https://doi.org/10.1063/5.0009698} {\bibfield  {journal}
  {\bibinfo  {journal} {Applied Physics Letters}\ }\textbf {\bibinfo {volume}
  {116}},\ \bibinfo {pages} {264002} (\bibinfo {year} {2020})}\BibitemShut
  {NoStop}%
\bibitem [{\citenamefont {Exarhos}\ \emph {et~al.}(2017)\citenamefont
  {Exarhos}, \citenamefont {Hopper}, \citenamefont {Grote}, \citenamefont
  {Alkauskas},\ and\ \citenamefont {Bassett}}]{Exarhos2017}%
  \BibitemOpen
  \bibfield  {author} {\bibinfo {author} {\bibfnamefont {A.~L.}\ \bibnamefont
  {Exarhos}}, \bibinfo {author} {\bibfnamefont {D.~A.}\ \bibnamefont {Hopper}},
  \bibinfo {author} {\bibfnamefont {R.~R.}\ \bibnamefont {Grote}}, \bibinfo
  {author} {\bibfnamefont {A.}~\bibnamefont {Alkauskas}},\ and\ \bibinfo
  {author} {\bibfnamefont {L.~C.}\ \bibnamefont {Bassett}},\ }\bibfield
  {title} {\bibinfo {title} {{Optical Signatures of Quantum Emitters in
  Suspended Hexagonal Boron Nitride}},\ }\href
  {https://doi.org/10.1021/acsnano.7b00665} {\bibfield  {journal} {\bibinfo
  {journal} {ACS Nano}\ }\textbf {\bibinfo {volume} {11}},\ \bibinfo {pages}
  {3328} (\bibinfo {year} {2017})}\BibitemShut {NoStop}%
\bibitem [{\citenamefont {Wilson}(1927)}]{Wilson1927}%
  \BibitemOpen
  \bibfield  {author} {\bibinfo {author} {\bibfnamefont {E.~B.}\ \bibnamefont
  {Wilson}},\ }\bibfield  {title} {\bibinfo {title} {{Probable Inference, the
  Law of Succession, and Statistical Inference}},\ }\href
  {https://doi.org/10.1080/01621459.1927.10502953} {\bibfield  {journal}
  {\bibinfo  {journal} {Journal of the American Statistical Association}\
  }\textbf {\bibinfo {volume} {22}},\ \bibinfo {pages} {209} (\bibinfo {year}
  {1927})}\BibitemShut {NoStop}%
\bibitem [{\citenamefont {McCloskey}\ \emph {et~al.}(2020)\citenamefont
  {McCloskey}, \citenamefont {Dontschuk}, \citenamefont {Broadway},
  \citenamefont {Nadarajah}, \citenamefont {Stacey}, \citenamefont {Tetienne},
  \citenamefont {Hollenberg}, \citenamefont {Prawer},\ and\ \citenamefont
  {Simpson}}]{McCloskey2020}%
  \BibitemOpen
  \bibfield  {author} {\bibinfo {author} {\bibfnamefont {D.~J.}\ \bibnamefont
  {McCloskey}}, \bibinfo {author} {\bibfnamefont {N.}~\bibnamefont
  {Dontschuk}}, \bibinfo {author} {\bibfnamefont {D.~A.}\ \bibnamefont
  {Broadway}}, \bibinfo {author} {\bibfnamefont {A.}~\bibnamefont {Nadarajah}},
  \bibinfo {author} {\bibfnamefont {A.}~\bibnamefont {Stacey}}, \bibinfo
  {author} {\bibfnamefont {J.~P.}\ \bibnamefont {Tetienne}}, \bibinfo {author}
  {\bibfnamefont {L.~C.}\ \bibnamefont {Hollenberg}}, \bibinfo {author}
  {\bibfnamefont {S.}~\bibnamefont {Prawer}},\ and\ \bibinfo {author}
  {\bibfnamefont {D.~A.}\ \bibnamefont {Simpson}},\ }\bibfield  {title}
  {\bibinfo {title} {{Enhanced Widefield Quantum Sensing with Nitrogen-Vacancy
  Ensembles Using Diamond Nanopillar Arrays}},\ }\href
  {https://doi.org/10.1021/acsami.9b19397} {\bibfield  {journal} {\bibinfo
  {journal} {ACS Applied Materials and Interfaces}\ }\textbf {\bibinfo {volume}
  {12}},\ \bibinfo {pages} {13421} (\bibinfo {year} {2020})}\BibitemShut
  {NoStop}%
\bibitem [{\citenamefont {Jaffe}\ \emph {et~al.}(2019)\citenamefont {Jaffe},
  \citenamefont {Felgen}, \citenamefont {Gal}, \citenamefont {Kornblum},
  \citenamefont {Reithmaier}, \citenamefont {Popov},\ and\ \citenamefont
  {Orenstein}}]{Jaffe2019}%
  \BibitemOpen
  \bibfield  {author} {\bibinfo {author} {\bibfnamefont {T.}~\bibnamefont
  {Jaffe}}, \bibinfo {author} {\bibfnamefont {N.}~\bibnamefont {Felgen}},
  \bibinfo {author} {\bibfnamefont {L.}~\bibnamefont {Gal}}, \bibinfo {author}
  {\bibfnamefont {L.}~\bibnamefont {Kornblum}}, \bibinfo {author}
  {\bibfnamefont {J.~P.}\ \bibnamefont {Reithmaier}}, \bibinfo {author}
  {\bibfnamefont {C.}~\bibnamefont {Popov}},\ and\ \bibinfo {author}
  {\bibfnamefont {M.}~\bibnamefont {Orenstein}},\ }\bibfield  {title} {\bibinfo
  {title} {{Deterministic Arrays of Epitaxially Grown Diamond Nanopyramids with
  Embedded Silicon-Vacancy Centers}},\ }\href
  {https://doi.org/10.1002/adom.201800715} {\bibfield  {journal} {\bibinfo
  {journal} {Advanced Optical Materials}\ }\textbf {\bibinfo {volume} {7}},\
  \bibinfo {pages} {1800715} (\bibinfo {year} {2019})}\BibitemShut {NoStop}%
\bibitem [{\citenamefont {Bradac}\ \emph {et~al.}(2010)\citenamefont {Bradac},
  \citenamefont {Gaebel}, \citenamefont {Naidoo}, \citenamefont {Sellars},
  \citenamefont {Twamley}, \citenamefont {Brown}, \citenamefont {Barnard},
  \citenamefont {Plakhotnik}, \citenamefont {Zvyagin},\ and\ \citenamefont
  {Rabeau}}]{Bradac2010}%
  \BibitemOpen
  \bibfield  {author} {\bibinfo {author} {\bibfnamefont {C.}~\bibnamefont
  {Bradac}}, \bibinfo {author} {\bibfnamefont {T.}~\bibnamefont {Gaebel}},
  \bibinfo {author} {\bibfnamefont {N.}~\bibnamefont {Naidoo}}, \bibinfo
  {author} {\bibfnamefont {M.~J.}\ \bibnamefont {Sellars}}, \bibinfo {author}
  {\bibfnamefont {J.}~\bibnamefont {Twamley}}, \bibinfo {author} {\bibfnamefont
  {L.~J.}\ \bibnamefont {Brown}}, \bibinfo {author} {\bibfnamefont {A.~S.}\
  \bibnamefont {Barnard}}, \bibinfo {author} {\bibfnamefont {T.}~\bibnamefont
  {Plakhotnik}}, \bibinfo {author} {\bibfnamefont {A.~V.}\ \bibnamefont
  {Zvyagin}},\ and\ \bibinfo {author} {\bibfnamefont {J.~R.}\ \bibnamefont
  {Rabeau}},\ }\bibfield  {title} {\bibinfo {title} {{Observation and control
  of blinking nitrogen-vacancy centres in discrete nanodiamonds}},\ }\href
  {https://doi.org/10.1038/nnano.2010.56} {\bibfield  {journal} {\bibinfo
  {journal} {Nature Nanotechnology}\ }\textbf {\bibinfo {volume} {5}},\
  \bibinfo {pages} {345} (\bibinfo {year} {2010})}\BibitemShut {NoStop}%
\bibitem [{\citenamefont {Nirmal}\ \emph {et~al.}(1996)\citenamefont {Nirmal},
  \citenamefont {Dabbousi}, \citenamefont {Bawendi}, \citenamefont {Macklin},
  \citenamefont {Trautman}, \citenamefont {Harris},\ and\ \citenamefont
  {Brus}}]{Nirmal1996}%
  \BibitemOpen
  \bibfield  {author} {\bibinfo {author} {\bibfnamefont {M.}~\bibnamefont
  {Nirmal}}, \bibinfo {author} {\bibfnamefont {B.~O.}\ \bibnamefont
  {Dabbousi}}, \bibinfo {author} {\bibfnamefont {M.~G.}\ \bibnamefont
  {Bawendi}}, \bibinfo {author} {\bibfnamefont {J.~J.}\ \bibnamefont
  {Macklin}}, \bibinfo {author} {\bibfnamefont {J.~K.}\ \bibnamefont
  {Trautman}}, \bibinfo {author} {\bibfnamefont {T.~D.}\ \bibnamefont
  {Harris}},\ and\ \bibinfo {author} {\bibfnamefont {L.~E.}\ \bibnamefont
  {Brus}},\ }\bibfield  {title} {\bibinfo {title} {{Fluorescence intermittency
  in single cadmium selenide nanocrystals}},\ }\href
  {https://doi.org/10.1038/383802a0} {\bibfield  {journal} {\bibinfo  {journal}
  {Nature}\ }\textbf {\bibinfo {volume} {383}},\ \bibinfo {pages} {802}
  (\bibinfo {year} {1996})}\BibitemShut {NoStop}%
\bibitem [{\citenamefont {Davan{\c{c}}o}\ \emph {et~al.}(2014)\citenamefont
  {Davan{\c{c}}o}, \citenamefont {Hellberg}, \citenamefont {Ates},
  \citenamefont {Badolato},\ and\ \citenamefont {Srinivasan}}]{Davanco2014}%
  \BibitemOpen
  \bibfield  {author} {\bibinfo {author} {\bibfnamefont {M.}~\bibnamefont
  {Davan{\c{c}}o}}, \bibinfo {author} {\bibfnamefont {C.~S.}\ \bibnamefont
  {Hellberg}}, \bibinfo {author} {\bibfnamefont {S.}~\bibnamefont {Ates}},
  \bibinfo {author} {\bibfnamefont {A.}~\bibnamefont {Badolato}},\ and\
  \bibinfo {author} {\bibfnamefont {K.}~\bibnamefont {Srinivasan}},\ }\bibfield
   {title} {\bibinfo {title} {{Multiple time scale blinking in InAs quantum dot
  single-photon sources}},\ }\href {https://doi.org/10.1103/PhysRevB.89.161303}
  {\bibfield  {journal} {\bibinfo  {journal} {Phys. Rev. B}\ }\textbf {\bibinfo
  {volume} {89}},\ \bibinfo {pages} {161303} (\bibinfo {year}
  {2014})}\BibitemShut {NoStop}%
\end{thebibliography}%

\end{document}